\documentclass[submission,Phys]{SciPost}
\usepackage{color,amsmath,amssymb,multirow,booktabs,graphicx,mathtools} 
\usepackage[T1]{fontenc}
\usepackage{hyperref}

\usepackage{float}

\newcommand{\arxiv}[1]{\href{http://arxiv.org/abs/#1}{arXiv:#1}}

\newcommand\one{\leavevmode\hbox{\small1\normalsize\kern-.33em1}}

\newcommand{\qqqquad}{\qquad \qquad \qquad}

\newcommand{\gev}{\text{GeV}}

\def\slashchar#1{\setbox0=\hbox{$#1$}           
   \dimen0=\wd0                                 
   \setbox1=\hbox{/} \dimen1=\wd1               
   \ifdim\dimen0>\dimen1                        
      \rlap{\hbox to \dimen0{\hfil/\hfil}}      
      #1                                        
   \else                                        
      \rlap{\hbox to \dimen1{\hfil$#1$\hfil}}   
      /                                         
   \fi}

\newcommand{\eg}{\textsl{e.g.}\;}

\newcommand{\be}{\begin{eqnarray*}}
\newcommand{\ee}{\end{eqnarray*}}

\newcommand{\bee}{\begin{eqnarray}}
\newcommand{\eee}{\end{eqnarray}}
\newcommand{\beeq}{\begin{equation}}
\newcommand{\eeeq}{\end{equation}}

\DeclareMathOperator\erf{erf}

\newcommand\Tstrut{\rule{0pt}{2.5ex}}         

\newcommand{\kwak}{\textsc{Kwak}}

\begin{document}

\begin{center}{\Large \textbf{
Multi-scale Mining of Kinematic Distributions with Wavelets
}}\end{center}

\begin{center}
Ben G. Lillard\textsuperscript{1}, 
Tilman Plehn\textsuperscript{2},
Alexis Romero\textsuperscript{1}, and
Tim M. P. Tait\textsuperscript{1}
\end{center}

\begin{center}
{\bf 1} Department of Physics and Astronomy, University of California, Irvine, USA \\
{\bf 2} Institut f\"ur Theoretische Physik, Universit\"at Heidelberg, Germany
\end{center}



\section*{Abstract}
{\bf Typical LHC analyses search for local features 
in kinematic distributions. 
Assumptions about anomalous patterns limit them to a relatively narrow
subset of possible signals.  Wavelets extract information from an
entire distribution and decompose it at all scales, simultaneously
searching for features over a wide range of scales.  We propose a
systematic wavelet analysis and show how bumps, bump-dip combinations,
and oscillatory patterns are extracted.  Our kinematic wavelet
analysis kit KWAK provides a publicly available framework to analyze
and visualize general distributions.}

\vspace{10pt}

\noindent\rule{\textwidth}{1pt}
\tableofcontents\thispagestyle{fancy}
\noindent\rule{\textwidth}{1pt}

\vspace{\fill}

\begin{flushright}
UCI-TR-2019-18
\end{flushright}

\clearpage
\section{New Physics at Multiple Scales}
\label{sec:intro}

Despite the proliferation of advanced statistical methods at the LHC,
simple analysis of well-chosen kinematic distributions remain a
powerful first attempt to tease out new physics with fuzzily specified
characteristics.  Resonances in invariant mass distributions or
enhanced tails at high energies can reveal the existence of new
particles produced on-shell, or the presence of heavy physics manifest
as higher-dimensional operators, respectively.

Simple analyses are also particularly amenable to data-driven
background determination.  For example, a resonance search in an
invariant mass distribution relies on a sideband fit, leading to a
background-only hypothesis given as a simple functional form.
At any point along the invariant mass distribution the analysis
searches for an excess or bump via a sliding mass window. The
underlying assumption is that the signal is a local excess, so the
window is characterized by a scale related to the resonance width.
This is also the origin of the look-elsewhere effect, which links the local
significance to a global significance based on treating
the entire distribution as one measurement.

The situation becomes more complicated when we search for more generic
patterns.  For example, quantum interference between the resonant
signal and the smooth background typically implies that the deviation
from the background becomes a deficit together with the excess, or a
bump-dip~\cite{dipbump_higgs,charlotte,Carena:2004xs}.
It is particularly prominent when the resonant particle has a large
width.  A typical bump hunt combines the bump-dip to a net excess,
considerably weakening the search.

There exist new physics models where modifications to the background
are even less localized.  Theories with compact extra
dimensions~\cite{Kaluza:1921tu} and their 4D product gauge
group~\cite{ArkaniHamed:2001ca} or clockwork~\cite{clockwork}
analogues predict towers of states, implying periodic invariant mass
patterns.  While individual resonant structures are local and amenable
to searches for bumps, an optimal search requires us to consider the
entire distribution.\medskip

The general question for analyses of a single kinematic distributions
is whether there exists an approach which balances the power of
searching for local features with the flexibility of searches which
retain information about longer scales or global features.  Wavelet
transforms are a standard tool which simultaneously decomposes data on
an interval into different scales, allowing for sensitivity to local
and global features.  The wavelet transform
\begin{enumerate}
\setlength\itemsep{-0.3em}
\item retains all information from the distribution in an orthogonal decomposition basis;
\item automatically zooms in to the proper resolution to match a given anomaly; and
\item retains all of the local information about the features of the distribution.
\end{enumerate}
Wavelets have been successfully applied to a number of analyses in
particle
physics~\cite{wavelet_tim,Cogan:2014oua,wavelet_monk,wavelet_gammaray}.
Applied to kinematic LHC data, they systematically evaluate the
complete kinematic distribution, without any assumptions about the
shape or scale of the potential anomaly.  Because they represent an
orthogonal change of basis, they maps the contents of a given number
of bins onto the same number of wavelet coefficients, allowing us to
mine a distribution for new physics without loss of
information.\medskip

In this short paper we introduce the Haar wavelet transform as a tool
to search for new physics in a kinematic LHC distribution. We
introduce the Haar wavelet and illustrate its main features in
Sec.~\ref{sec:wavelets}, considering idealized deviations in the form
of narrow and broad bumps, bump-dips, and an oscillatory pattern.  In
Sec.~\ref{sec:maa} we apply our analysis to simulated data inspired by
the ATLAS di-photon invariant mass~\cite{atlas}, injecting the same
set of signal patterns.  We analyze the actual ATLAS di-photon
distribution in Sec.~\ref{sec:atlas}.  Appendices include some details
of the statistical analysis, and introduce our publicly available
Python analysis package, \kwak.

\section{Wavelet Transform} 
\label{sec:wavelets}

A Wavelet transform represents a given function in terms of simple
orthonormal basis.  In that sense it is similar to a Fourier
transform, with the main difference that the wavelet basis retains a
notion of locality in position space, which is relinquished by the
Fourier transform. 

\subsection{Haar wavelet}

A particularly simple wavelet is the Haar wavelet in one
dimension~\cite{haar}, defined on the interval $x \in [0,1]$. The
first two basis functions are
\begin{align}
h_0(x) = 1  \qquad \text{and} \qquad 
h_1(x) = \begin{cases} +1 & x = 0~...~1/2 \\
                       -1 & x = 1/2~...~1 \; .
         \end{cases} 
\end{align}
They characterize the over-all normalization of the function and its
relative change from one side of the interval to the other,
respectively.  The next two basis functions are constructed from
$h_1(x)$, compressed in $x$ by a factor of two,
\begin{align}
h_{2,1}(x) = \sqrt{2} \; h_1(2x)
\qqqquad 
h_{2,2}(x) = \sqrt{2} \; h_1(2x-1) \; .
\end{align}
They characterize the change from one side of each subintervals to
the other.  Further basis functions continue to subdivide the
intervals from the previous level.  For example, the next step defines
four functions, compressed by an additional factor two,
\begin{align}
h_{3,1}(x) &= 2 \; h_1(4x) 
\qquad 
&h_{3,2}(x) &= 2 \; h_1(4x-1) 
\notag \\
h_{3,3}(x) &= 2 \; h_1(4x-2) 
\qquad 
&h_{3,4}(x) &= 2 \; h_1(4x-3) \; .
\end{align}
Continuing to sub-divide the $x$-interval, the higher wavelet
functions $h_{\ell,m}$ are organized in families labelled by level
$\ell$ and increasingly localized in $x$. The label
$m=1~...~2^{\ell-1}$ specifies their position inside the interval.
With the normalization $h_{\ell m} \propto 2^{(\ell-1)/2}$ the
real wavelet functions are orthonormal,
\begin{align}
\int_0^1 dx \; h_{\ell,m}(x) ~h_{\ell',m'}(x) = \delta_{\ell \ell'} \; \delta_{mm'} \; ,
\end{align}
allowing the wavelet representation of a function $f(x)$ to be easily inverted,
\begin{align}
f(x) = \sum_{\ell,m} \tilde{f}_{\ell,m} \; h_{\ell m}(x) 
\qquad \Leftrightarrow \qquad 
\tilde{f}_{\ell,m} = \int_0^1 dx \; h_{\ell,m}(x) f(x) \; .
\label{eq:wavelet_trafo}
\end{align}
In this notation the similarity to a Fourier transform is manifest:
the wavelets at each level resolve a waveform pattern that is the
$\ell$th harmonic of the interval, but divided into $2^{\ell-1}$
locations along the interval, saturating the Nyquist criterion.  The
first coefficient $\tilde{f}_0$ is special in that it represents the
over-all normalization of the distribution, and we will neglect it in
most of our shape analysis below.\medskip

\newpage

\begin{figure}[H]
\centering
\def\quadwid{0.55}
\def\midwid{-0.025}
\def\leftwid{-0.017}
\hspace*{\leftwid \textwidth}
\begin{tabular}{r r}
\includegraphics[height=\quadwid \textwidth]{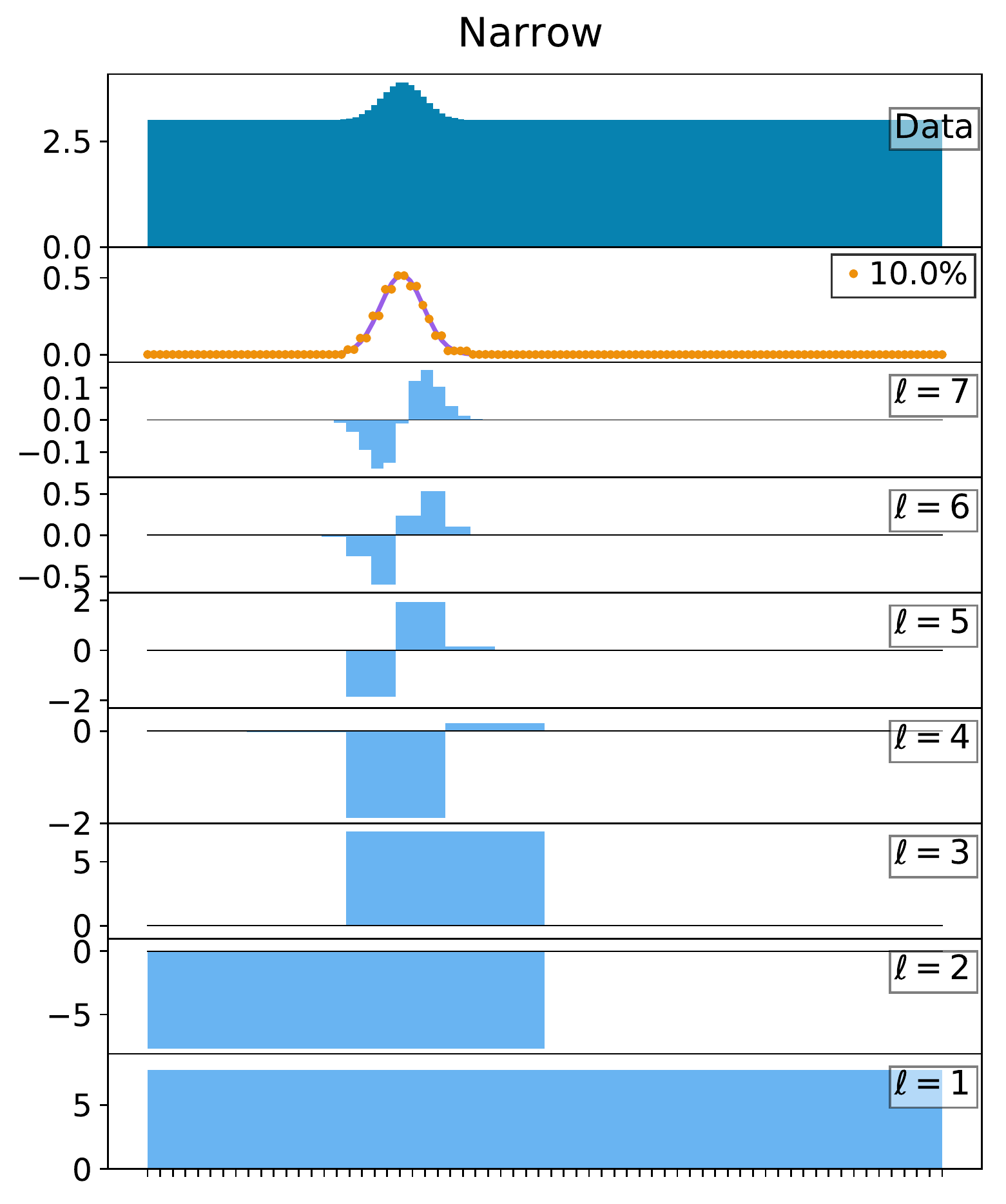} &
\hspace*{\midwid \textwidth}
\includegraphics[height=\quadwid \textwidth]{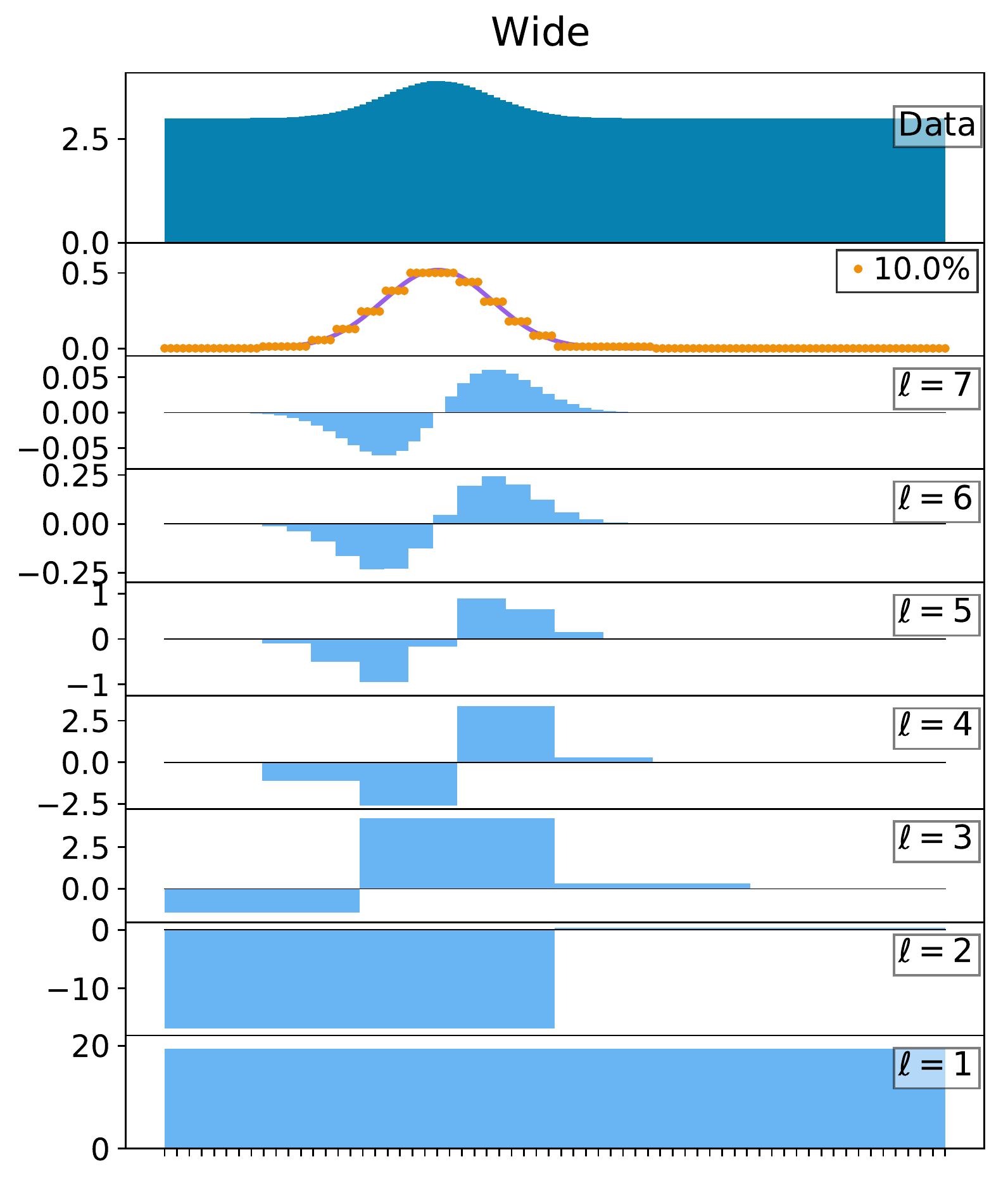}  
\\ 
\hspace*{\leftwid \textwidth}
\includegraphics[height=\quadwid \textwidth]{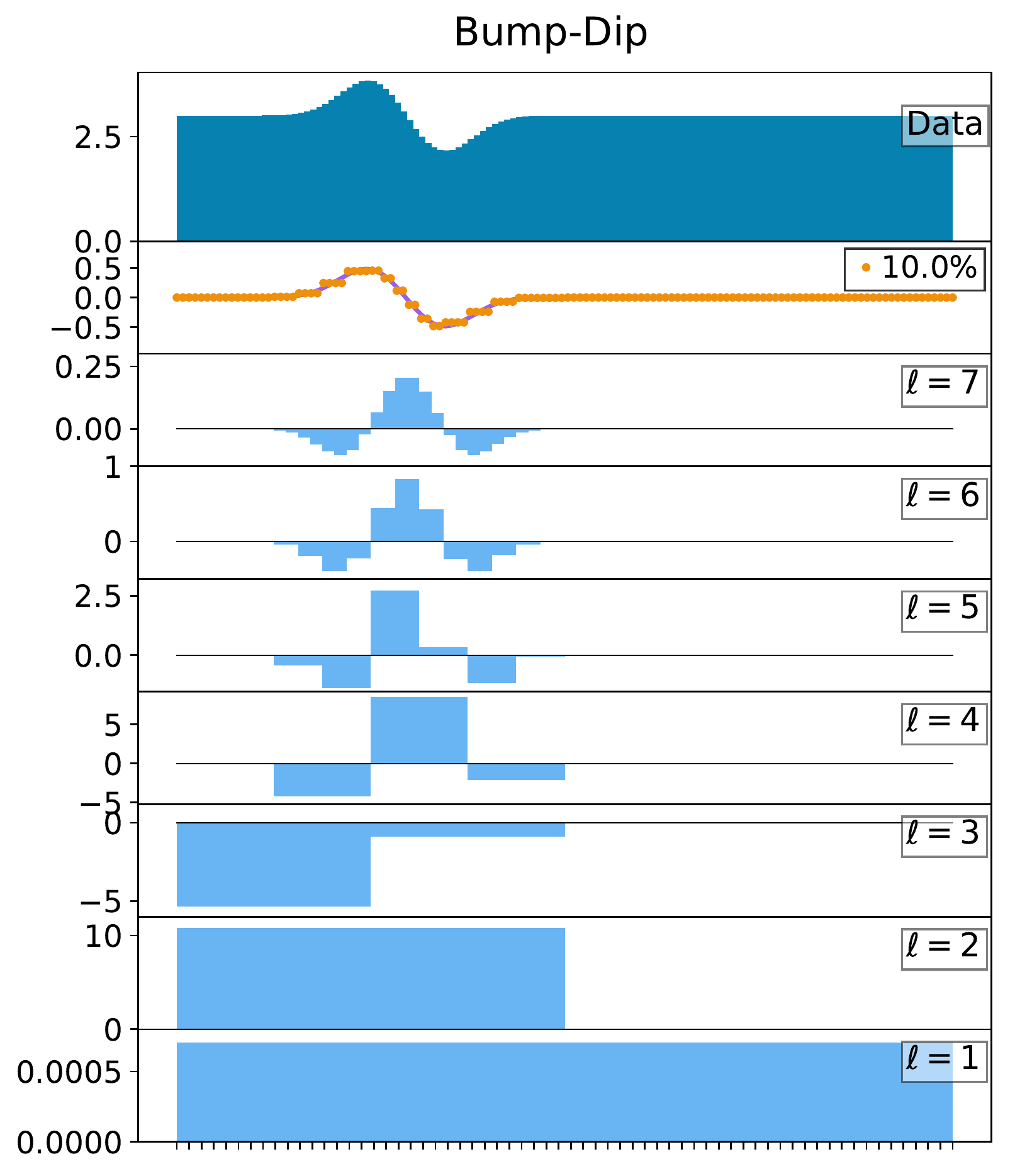} &
\hspace*{\midwid \textwidth}
\includegraphics[height=\quadwid \textwidth]{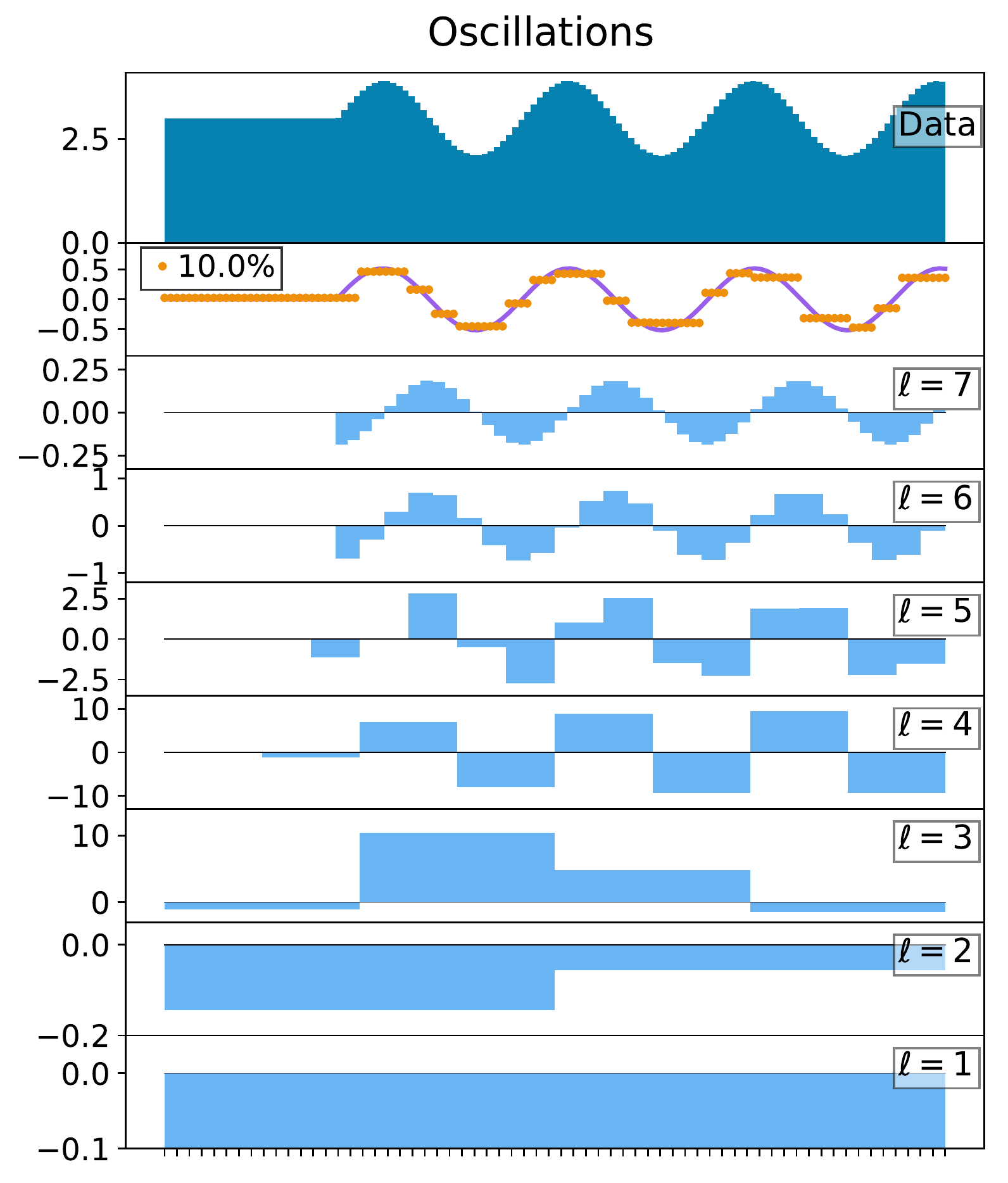}
\end{tabular}
\caption{Toy wavelet analysis for a narrow (upper left) and a wide
  (upper right) bump, a bump-dip (lower left), and an oscillatory
  signal (lower right) on top of a flat background.  The top panel
  shows the original distribution, the one below the pattern
  reconstructed retaining the largest $10\%$ wavelet coefficients, and
  the remaining panels show the values of the wavelet coefficients $\tilde{f}_1$
  through $\tilde{f}_{7,m}$ for 128 bins and no statistical
  fluctuations. The reconstructed signal (orange) is overlaid on top of the original (in purple).
   For each level the coefficients are aligned with their
  actual position in the distribution.}
\label{fig:nostats}
\end{figure}

A kinematic distribution $f(x)$ with $2^L$ bins $f_j$ defines $L$
levels of wavelet coefficients. Including $\tilde{f}_0$, there are a
total of $2^L$ wavelet coefficients, and the wavelet coefficients
contain precisely the same information as the number of bin in the
distribution.  Because each wavelet basis state spans two distinct
regions, the resolution at level $\ell$ corresponds to $2 \times
2^{\ell -1} = 2^\ell$ bins.  From the definition of the wavelet
transform in Eq.\eqref{eq:wavelet_trafo} it is clear that, 
for example, the highest wavelet coefficients encode the $2^L/2$ pairwise
differences between neighboring bins,
\begin{align}
\tilde{f}_{L,m} =  f_{2m-1} - f_{2m} 
\qquad \text{for $m=1~...~2^{L-1}$,}
\label{eq:wavelet_m}
\end{align}
where in the discretized distribution $f(x) = f_j$ the bin index $j = 1 \ldots 2^L$ replaces the continuous parameter $x$.
The localized wavelet coefficients are aligned with the original
distribution $f(x)$ such that at the highest level each wavelet
coefficient $\tilde{f}_{L,m}$ corresponds to two bins $f_{2m-1}$ and
$f_{2m}$, and the next level corresponds to four bins, etc.  In many
applications of the wavelet transformation it is standard to normalize
the wavelet coefficients by a factor of $2^{(\ell-1)/2}$, but in our
statistical analysis of integer-valued signals the definition in
Eq.\eqref{eq:wavelet_m} is more convenient.

\subsection{Toy Examples}
\label{sec:wavelets_toy}

In Fig.~\ref{fig:nostats} we show the set of wavelet coefficients at
each level for four toy distributions:
\begin{enumerate}
\setlength\itemsep{-0.3em}
\item a narrow Gaussian bump;
\item a wide Gaussian bump;
\item a bump-dip combination; and
\item an oscillatory pattern with a shifted
starting point.
\end{enumerate}
Each distribution is added to a flat background and represented by a
histogram with 128 bins.  For the flat background alone all wavelet
coefficients vanish by definition, Eq.\eqref{eq:wavelet_m}.  In each
pane, the top panel shows the original histogram, and the lower panels
show the wavelet coefficients from $\ell = 7$ to $\ell = 1$, followed
by $\tilde{f}_0$ in the bottom panel.  In this toy
illustration we neglect statistical fluctuations, so the wavelet
coefficients correspond perfectly to the source distribution. As
discussed above, we align the wavelet coefficients of each level
$\ell$ with the corresponding bins of the original distribution
$f(x)$.\medskip

The upper left panel of Fig.~\ref{fig:nostats} with the narrow bump
illustrates how the large wavelet coefficients are localized at the
position of the narrow excess. The largest wavelet coefficients appear
at level $\ell=5$, where the entire bump is covered by the two
coefficients $\tilde{f}_{5,7}$ and $\tilde{f}_{5,8}$. This information
encodes the fact that we are looking at a localized feature of size
$1/2^5 \simeq 0.03$ of the original range $x=0~...~1$. Interesting
features can be reconstructed by considering a subset of the leading
wavelet coefficients, which contain the most important information,
\begin{align}
f_\text{approx}(x) = \sum_\text{leading $\tilde{f}$} \tilde{f}_{\ell,m} \; h_{\ell m}(x) \; .
\label{eq:wavelet_approx}
\end{align}
By removing subleading coefficients, contributions of limited
statistical significance are excised, allowing for sharp and robust
image of the deviation from the background model.  The second line in
the upper left panel shows the result from the leading $10\%$ of
wavelet coefficients in size.  Indeed, the small set of leading
wavelets describe the bump pattern well, at the expense only of
resolution from the highest level, $\ell = 7$.  In the upper right
panel we repeat this analysis for a bump with twice the width. As
expected, most of the power is contained in the $\ell=4$ coefficients.

The lower left panel of Fig.~\ref{fig:nostats} describes a bump-dip,
as it for example appears through quantum interference with wide
resonances~\cite{charlotte}. It is a challenge to the standard
bump-hunting methods, which average the bump and the dip structures
unless the resolution is sufficient and very carefully tuned.  The
total width of the feature is chosen to be about twice the width of
the narrow bump, and indeed the largest wavelet coefficient is
$\tilde{f}_{4,3}$, corresponding to the correct scale and position.
At this scale, both the bump and the dip individually contribute
positively to the wavelet coefficient.

Finally, an off-set oscillatory pattern is assumed for the lower right
panel of Fig.~\ref{fig:nostats}. Such a modification poses a serious
challenge for LHC searches~\cite{clockwork}.  The frequency of the
pattern is such that most of its power appears at $\ell=4$ with $m >
2$, reflecting the fact that the oscillations begin after an initial
gap.  We also show the approximate reconstructed signal, retaining the
leading $10\%$ wavelet coefficients, confirming that the signal
pattern is again well described.

\begin{figure}[ht]
\includegraphics[width=0.47\textwidth]{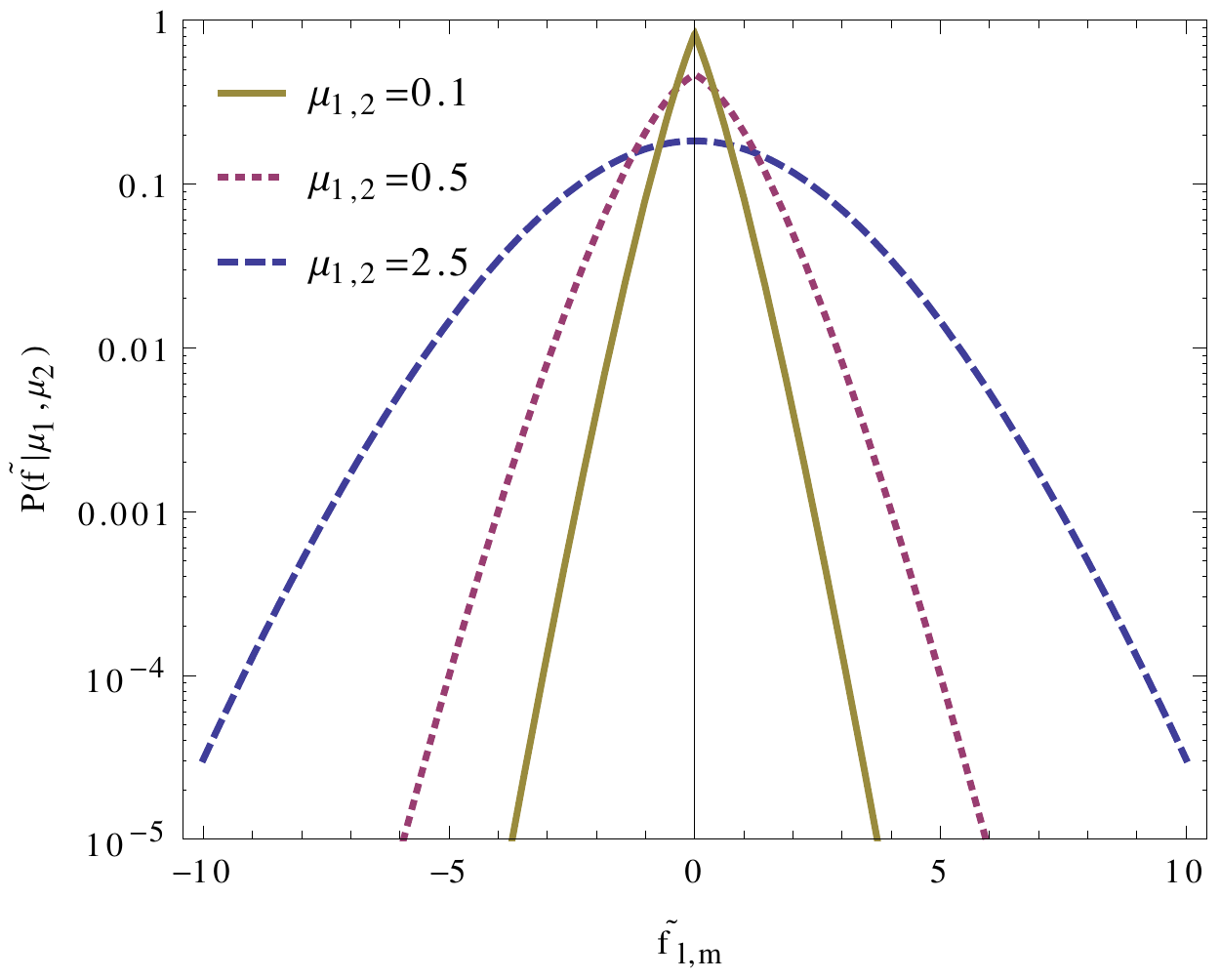}
\hspace*{0.05\textwidth}
\includegraphics[width=0.47\textwidth]{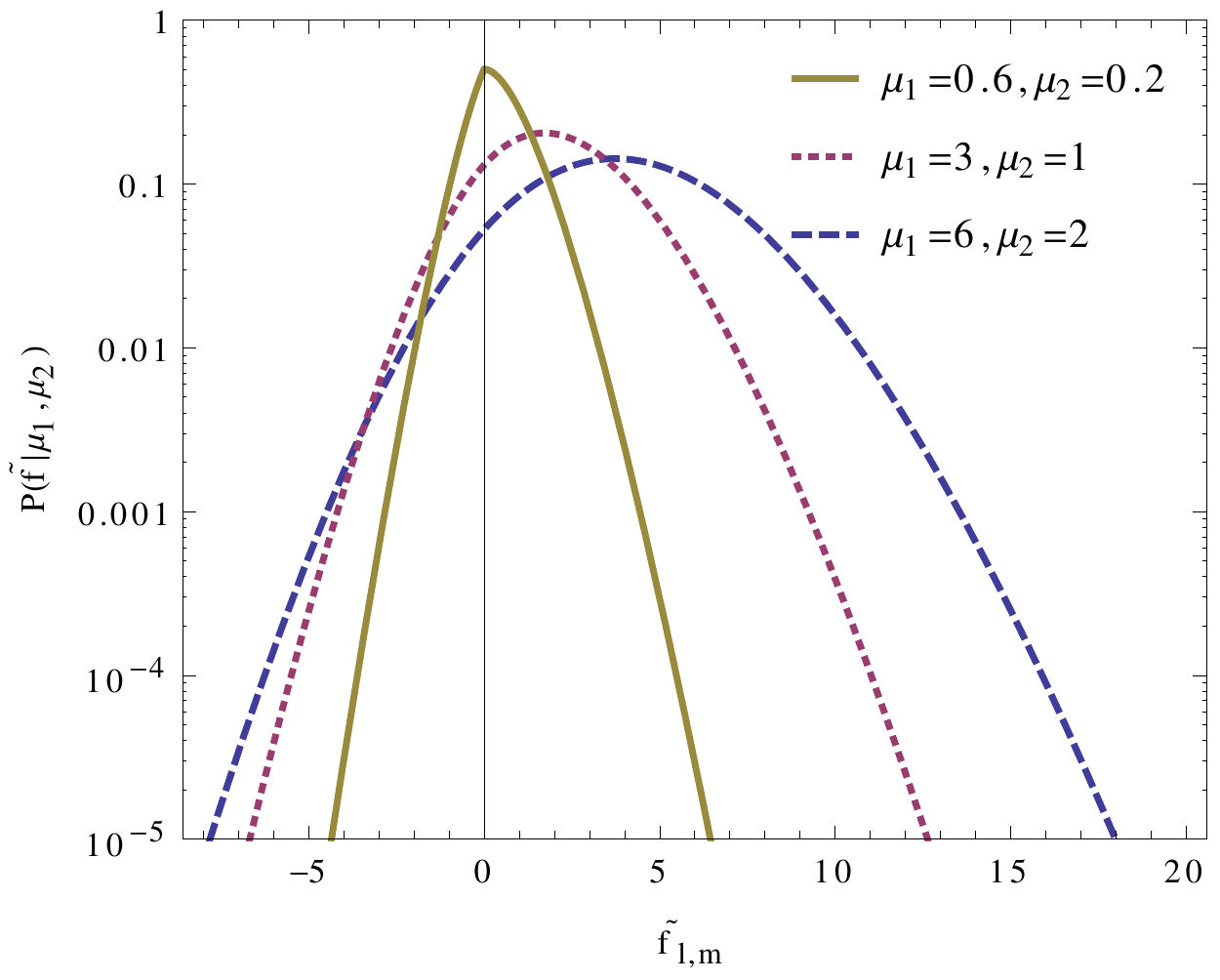} 
\vspace*{-6mm}
\caption{Statistical distribution for the wavelet coefficient
  $\tilde{f}$ assuming Poisson distributions of the two bins of the
  kinematic distribution $f_{1,2}$. The two input distributions are
  described by their means $\mu_{1,2}$.}
\label{fig:distris}
\end{figure}

\subsection{Statistical Analysis}

Realistic distributions inevitably contain statistical fluctuations.
A kinematic distribution $f(x)$ is experimentally represented by $2^L$
bins $f_j$, where $f_j$ is the number of events in the $j\text{th}$ bin and is integer-valued.
If we assume that the bins are
statistically independent, each bin count is described by a Poisson
distribution with mean $\mu_j$,
\begin{align}
P(f_j|\mu_j) = \frac{e^{-\mu_j} \; \mu_j^{f_j}}{f_j!} \; ,
\label{eq:poiss_f}
\end{align}
which implies that the probability distribution for the $m=1$ wavelet
coefficient of the highest level $\ell = L$ is
\begin{align}
P(\tilde{f}|\mu_1,\mu_2) 
= \sum_{f_1,f_2} \frac{e^{-\mu_1-\mu_2} \; \mu_1^{f_1} \mu_2^{f_2}}{f_1! f_2!} 
\Bigg|_{\tilde{f} = f_1 - f_2} 
= e^{-\mu_1-\mu_2} \; \left( \frac{\mu_1}{\mu_2} \right)^{\tilde{f}/2}
   \mathcal{I}_{\tilde{f}}(2 \sqrt{\mu_1 \mu_2})
   \; ,
\label{eq:poiss_h}
\end{align}
where $\mathcal{I}_n$ is the $n$th modified Bessel function of the
first kind.  This probability distribution is referred to as the
Skellam distribution~\cite{irwin}.  Its mean, variance, skew, and
excess kurtosis are
\begin{align}
\mu= & \mu_1 - \mu_2 \; ,
&\quad 
\sigma^2= & \mu_1 + \mu_2 \; , \notag \\
\gamma_1= & \frac{ \mu_1 - \mu_2}{(\mu_1 + \mu_2)^{3/2}} \; ,
&\quad
\gamma_2= & \frac{1}{\mu_1 + \mu_2} \; .
\label{eq:skellamd}
\end{align}
When the Poisson distributions per bin in Eq.\eqref{eq:poiss_f}
becomes Gaussian, $\mu_1 + \mu_2 \gg 1$, $\gamma_1$ and $\gamma_2$
vanish, and $P(\tilde{f})$ approaches the expected Gaussian shape.  We
show the probability distribution for the wavelet coefficients in
Fig.~\ref{fig:distris}, assuming independent Poisson distributions for
the bins of the underlying kinematic distribution.  The tails of
$P(\tilde{f})$ are exponentially suppressed, and as the mean values
$\mu_{1,2}$ of the input distributions increase, the resulting
$P(\tilde{f})$ indeed approaches a Gaussian.  In
Appendix~\ref{appx:tool}, we provide the probability distribution
$P(\tilde{f} | H_0)$ for generic values of $\ell \leq L$ and $m \geq
1$, and for a generic hypothesis pattern $H_0$.

A statistical analysis traces all of the correlations of the input
distribution $f(x)$ in terms of the bin values $f_j$ to the wavelet
coefficients $\tilde{f}_j$. If we do nothing other than transform from
the $f_j$ to the $\tilde{f}_j$, the two descriptions are equivalent.
The power in the wavelet analysis is in how the deviations are
reflected in a subset of the wavelets, which simultaneously analyze
different scales and can be filtered to enhance specific kinds of
searches.  For example, the oscillatory pattern largely lives in a set
of wavelet coefficients of a single given level $\ell$.

\paragraph{Fixed Resolution Global Significance:}
From Eq.\eqref{eq:wavelet_m}, it is clear that each bin of the
distribution only contributes linearly to a single wavelet
coefficient. If the individual bins are statistically independent, the
wavelet coefficients for a single level are also statistically
independent, allowing them to be trivially combined into a single
statistical analysis. 

A $p$~value can be calculated from Eq.\eqref{eq:poiss_h} for each wavelet coefficient $\tilde{f}_{\ell, m}$, and translated into a test statistic $q_{\ell, m}$ defined as
\begin{equation}
q_{\ell, m} = - 2 \ln p_{\ell, m}\,,
\end{equation}
which obeys a $\chi^2$ distribution with two degrees of freedom. For wavelet coefficients of fixed $\ell$ the $q_{\ell, m}$ can be summed together to create a combined test statistic $q_\ell$,
\begin{equation}
q_\ell = \sum_{m=1}^{k} q_{\ell, m}\,.
\end{equation}
If the $\tilde{f}_{\ell, m}$ are statistically independent then $q_\ell$ follows a $\chi^2$ distribution with $2k$ degrees of freedom,
meaning that the statistical fluctuation in the ensemble of wavelet coefficients sharing the same $\ell$ can be easily quantified. In Eq.\eqref{eq:incgamma} in Appendix~\ref{appx:tool} we show that $p_\ell$, the combined $p$-value for all $\tilde{f}_{\ell, m}$ of a given $\ell$, can be written in terms of an incomplete gamma function.

This metric is highly useful for identifying features in the data that are spread 
over multiple coefficients within the same level of the wavelet transformation, 
and we refer to it as the fixed resolution global significance (FRGS).
The situation is more subtle when an analysis requires combining multiple
levels into a single statistical analysis, for example when searching
for different local features of different scales.

\section{Di-photon Mass Distribution} 

For a more realistic illustration we rely on a measured ATLAS
di-photon invariant mass spectrum, $m_{\gamma \gamma}$~\cite{atlas}.
With its statistical fluctuations it allows us to perform a
semi-realistic wavelet analysis with different injected signals. We
choose the same patterns as in Sec.~\ref{sec:wavelets_toy}. After that
we analyze the actual ATLAS results in a desperate attempt to search
for new physics at the LHC.

\begin{figure}[hb] 
\centering
\includegraphics[width=0.6\textwidth]{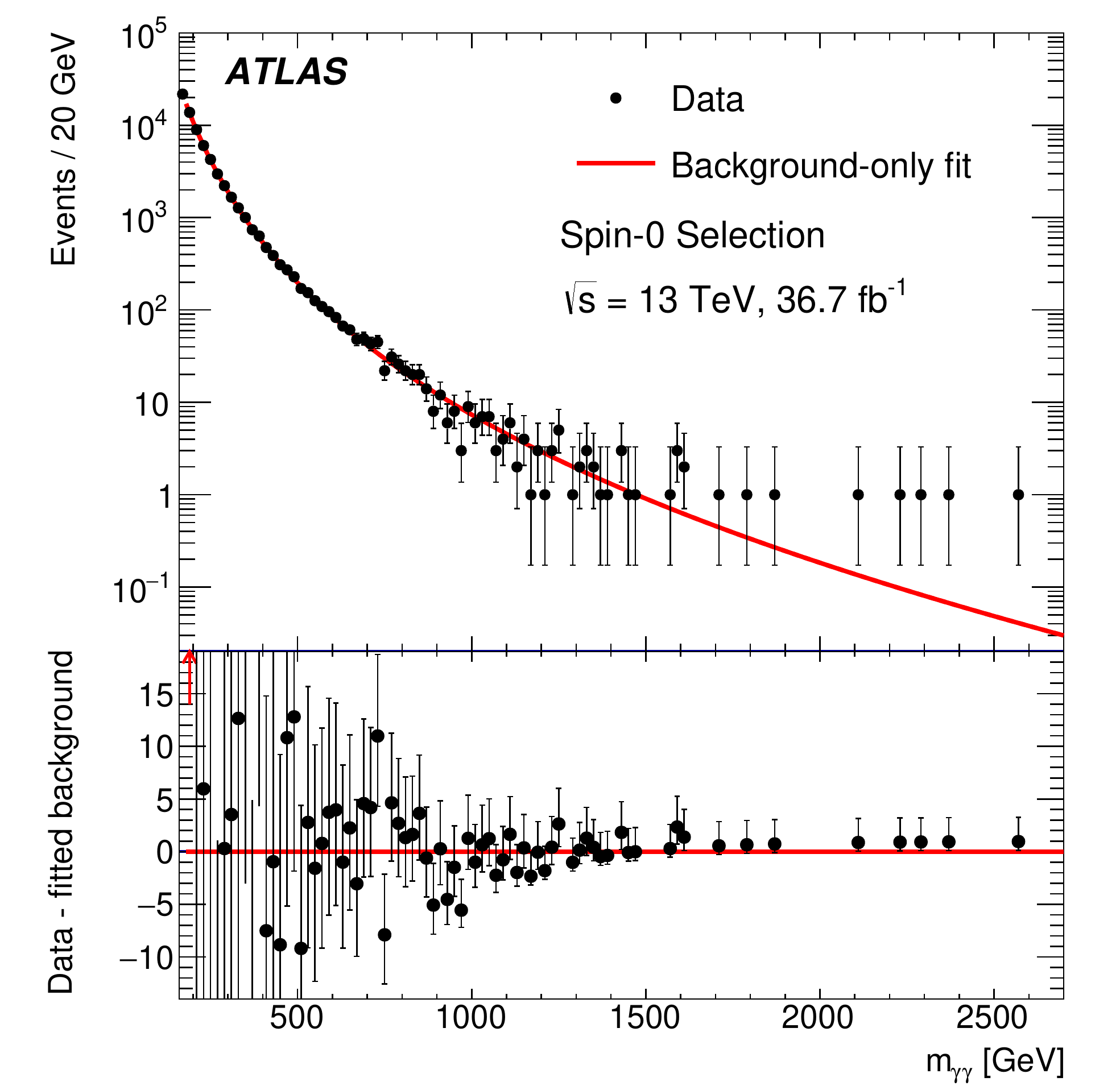}
\caption{Di-photon invariant mass distribution after spin-0 resonance
  search selection from ATLAS~\cite{atlas} and background-only fit
  (upper panel).  The lower panel shows the difference between data
  and the fit for each bin.}
\label{fig:atlas}
\end{figure}

\subsection{Injected Signals} 
\label{sec:maa}

\begin{figure}[ht]
\centering
\def\quadwid{0.52} 
\def\midwid{-0.025}
\def\leftwid{-0.02}
\hspace*{\leftwid \textwidth}
\begin{tabular}{r r}
\includegraphics[height=\quadwid \textwidth]{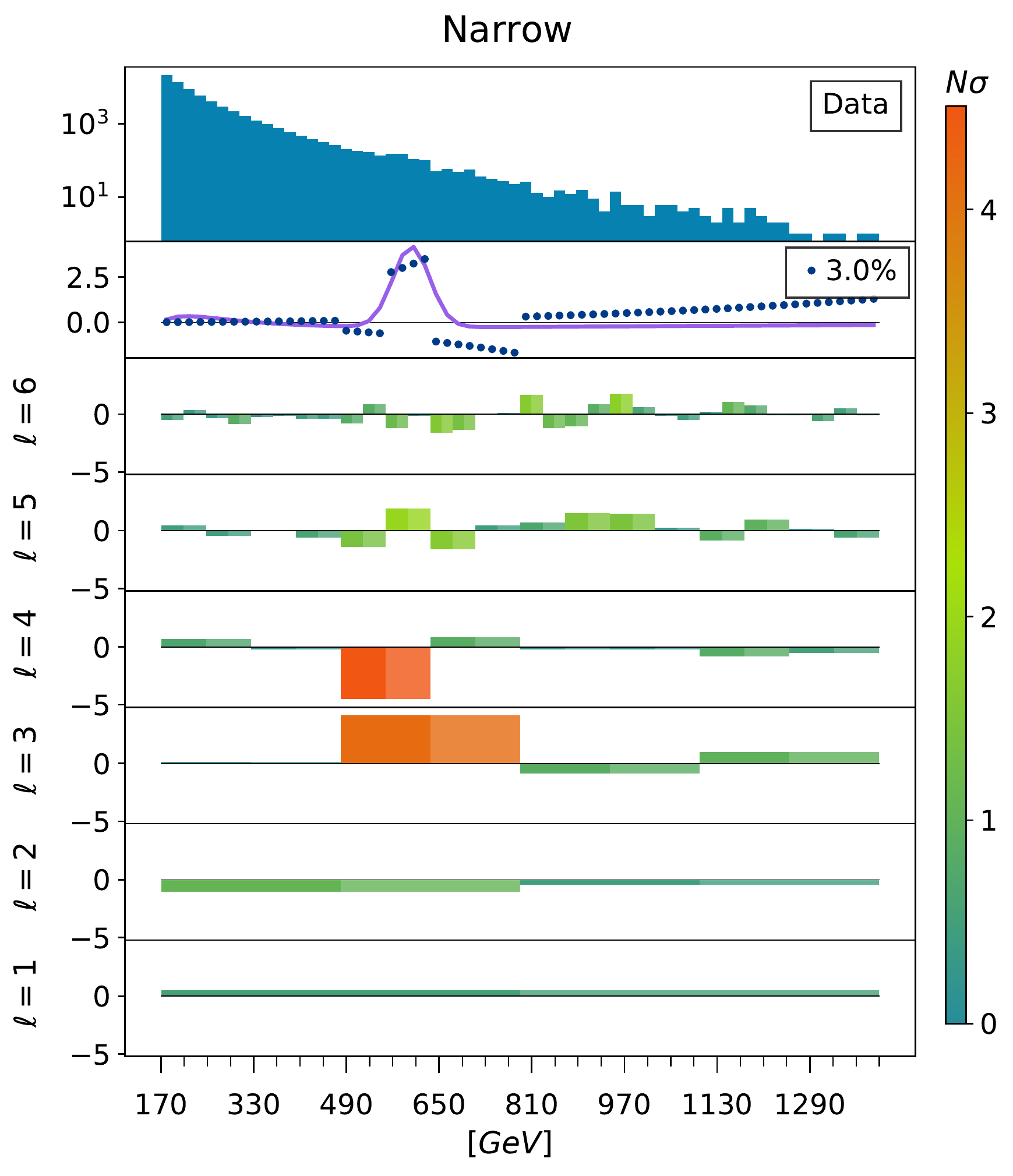} &
\hspace*{\midwid \textwidth}
\includegraphics[height=\quadwid \textwidth]{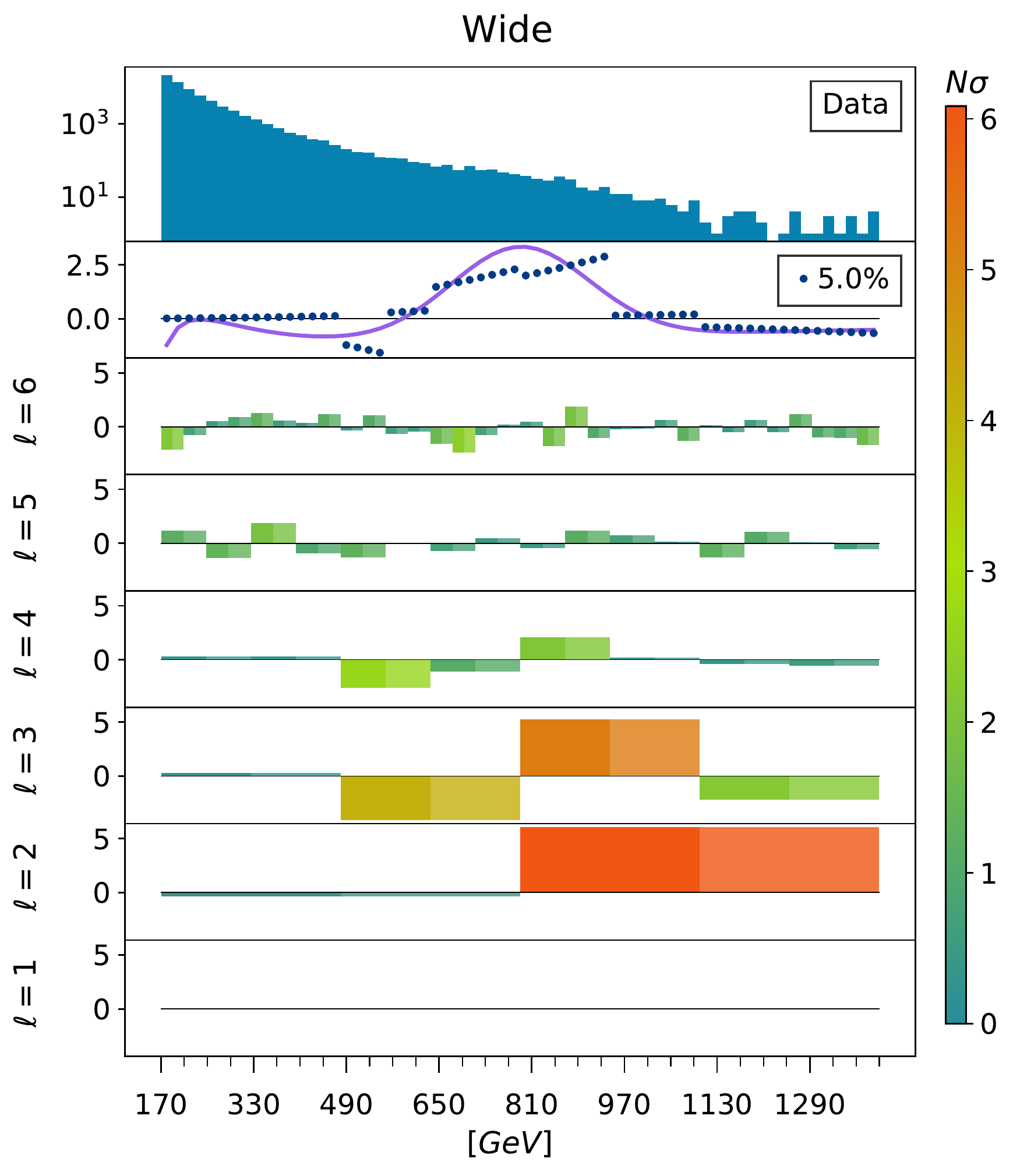}   
\\ 
\hspace*{\leftwid \textwidth}
\includegraphics[height=\quadwid \textwidth]{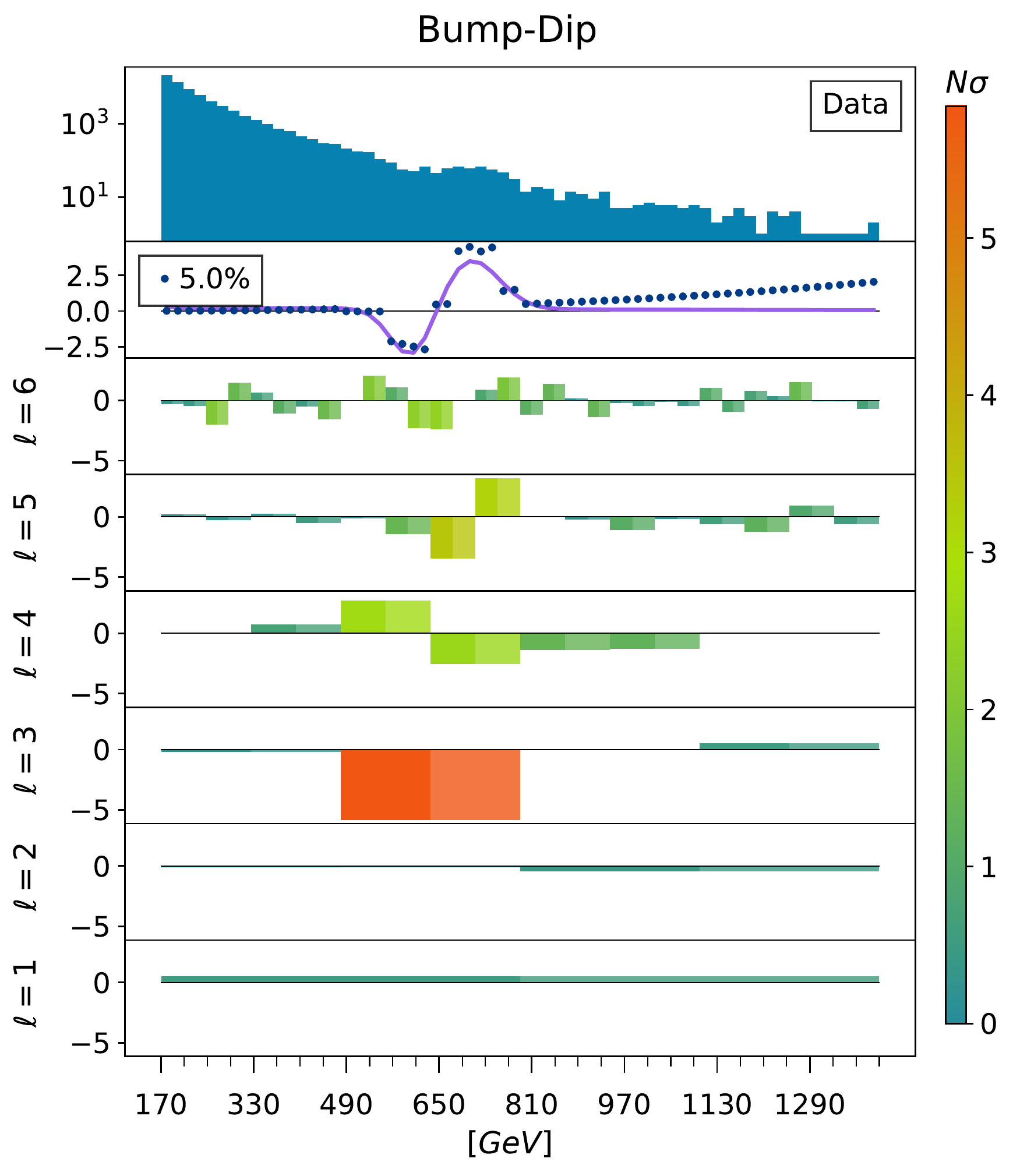} &
\hspace*{\midwid \textwidth}
\includegraphics[height=\quadwid \textwidth]{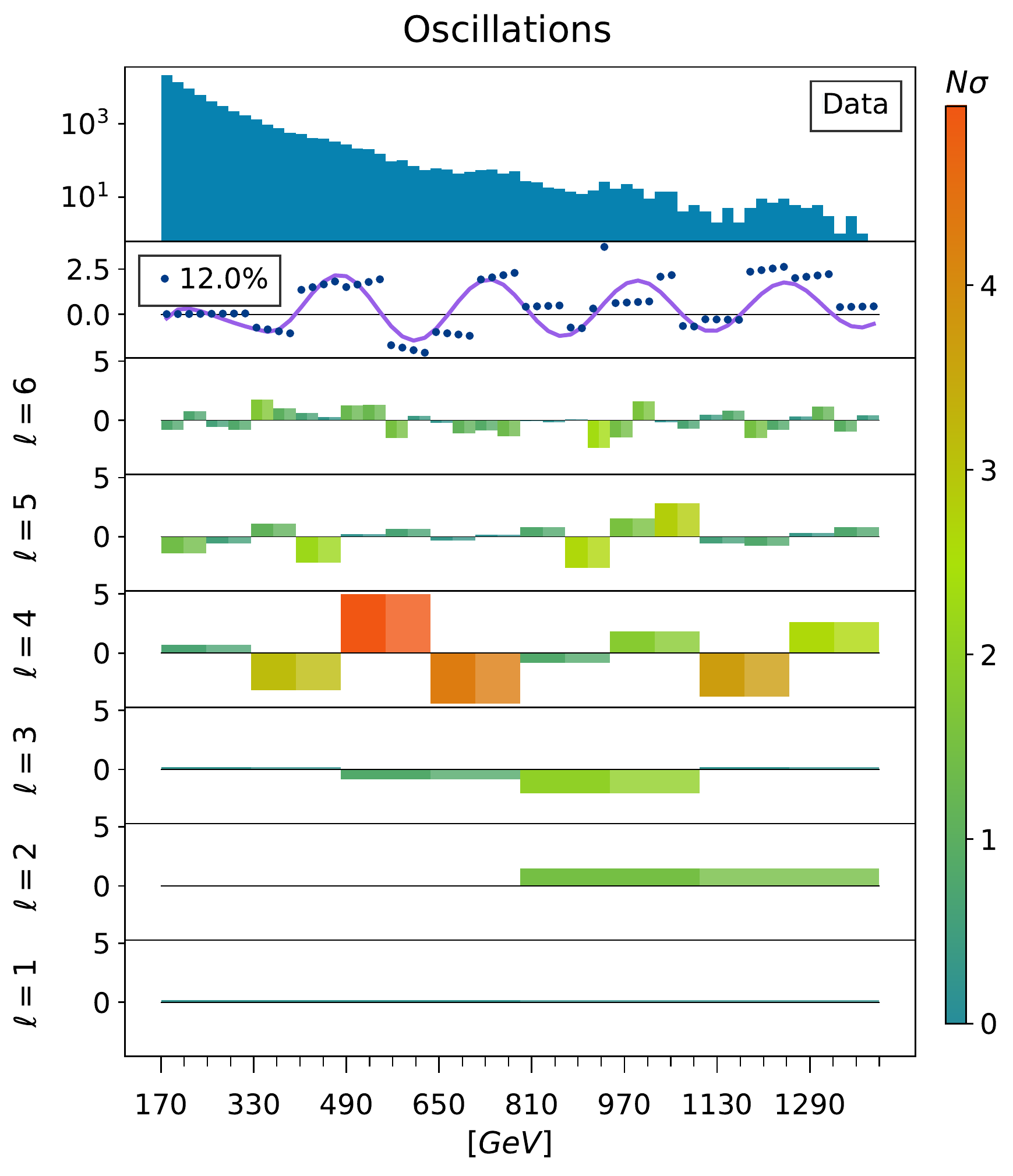}
\end{tabular}
\caption{Wavelet transform of the di-photon invariant mass
  distribution with background hypothesis fit to the ATLAS data. We
  inject a narrow resonance (top left), wide resonance (top right),
  bump-dip (lower left), and oscillation pattern (lower right).  The
  top panes show the input distribution, the next a signal
  reconstruction based on the indicated fraction of most significant
  coefficients, and the remaining panes the significance of each
  coefficient. The $x$-axis bins correspond to a linear scale between
  $m_{\gamma \gamma}=200$~GeV and 1.45~TeV. In the second panel
  we show the signal function (in purple) that was used to generate the data for 
  each function.
  Each wavelet coefficient is color-coded based on its deviation from 
  the background hypothesis, with a color scale chosen individually 
  for each plot based on the size of the most significant excess.}
\label{fig:examples}
\end{figure}

The background-only hypothesis for the ATLAS measurement shown in
Fig.~\ref{fig:atlas} is described by the functional
form~\cite{Aaboud:2016tru}
\begin{align}
f_B (x) = N ~(1 - x^{1/3})^b ~x^a
\qquad \text{with} \quad x = \frac{m_{\gamma \gamma}}{\sqrt{s}}
\label{eq:Hb}
\end{align}
We fit the coefficients $N$, $a$, and $b$ to the ATLAS di-photon
spectrum~\cite{atlas}, shown for reference in Fig.~\ref{fig:atlas},
and use this as a more realistic bases to inject the same four signal
patterns used before, namely
\begin{enumerate}
\setlength\itemsep{-0.3em}
\item a narrow Gaussian bump with mass 600~GeV and width 80~GeV;
\item a wide Gaussian bump with mass 750~GeV and width 300~GeV;
\item a bump-dip with a peak at 700~GeV and a dip 100~GeV below; and
\item an oscillation with a wave length of 265~GeV and a first peak at 415~GeV.
\end{enumerate}
The combined kinematic distribution is binned into a histogram,
subject to Poisson fluctuations. The injected signal pattern is
normalized to give an approximately $5\sigma$ deviation in at least
one of the wavelet coefficients.


The wavelet decompositions of the four resulting distributions are
shown in Fig.~\ref{fig:examples}.  The top pane of each panel shows
the resulting distribution in $m_{\gamma \gamma}$.  The lowest six
panes of each panel indicate the number of standard deviations in the
corresponding wavelet coefficient compared to the background-only
hypothesis, with color coding to guide the eye to more significant
deviations. The second pane of each panel shows
the reconstructed signal based on the indicated fraction of wavelet
coefficients most significantly different from the background.\medskip

From Fig.~\ref{fig:examples}, it is evident that both the narrow and
wide resonant examples show the power of the wavelet transform to pick
out the location and size of such a feature without making specific
analysis choices beyond the initial binning of the histogram. Both are
relatively well reconstructed with modest pixelation by a small
fraction of $3\%$ and $5\%$ of the most significantly deviating
wavelet coefficients.  As in the toy example, the bump-dip is much
more easily teased out by the wavelet that best matches its structure
than a typical resonance search would be able to handle.  In this
case, a $5.5\sigma$ deviation in the $\ell=3$, $m=2$ wavelet
coefficient correctly identifies its location and structure, and the
reconstruction based on the $5\%$ most significant wavelets reflects
its structure.  The oscillatory pattern is correctly identified at $\ell =4$, 
where the wavelet structure most closely matches the injected frequency.
Its reconstruction in the second pane of the plot reflects the
challenge of striking a balance between keeping enough coefficients to
faithfully reconstruct the wave form, while excluding statistical
noise and background.

As the reconstructed signal provides primarily qualitative information 
about the nature of the statistical excess, there is no ``correct'' number
of wavelet coefficients to use in the signal reconstruction. Instead of
keeping a particular fraction of the coefficients, one could just as easily
specify a minimum value of $N_\sigma$. Our choices in Fig.~\ref{fig:examples} 
to use $3\%$, $5\%$ or $12\%$ of the coefficients are roughly equivalent to
setting $N_\sigma^\text{min} \sim 2$.
Without relying on this subjective benchmark, the presence or 
absence of new physics can be inferred directly from the analysis of 
 individual wavelet coefficients, and from combined metrics like the 
fixed resolution global significance (FRGS).

\begin{figure}[ht]
\begin{center}
$\vcenter{\hbox{\includegraphics[width=0.6\textwidth]{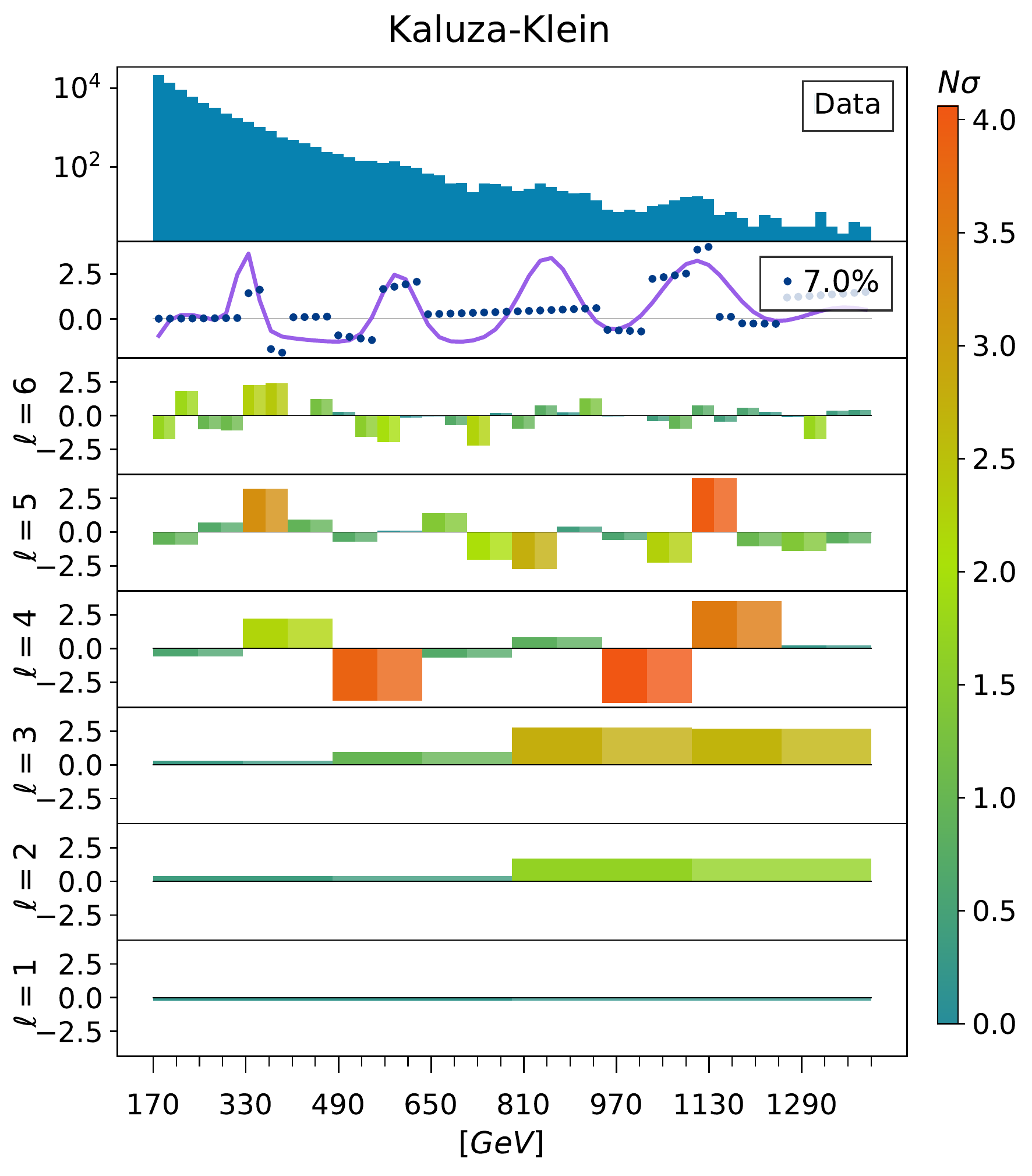}}}$
\hspace*{0.03\textwidth}
$\vcenter{\hbox{
\begin{tabular}{c|c } \toprule \Tstrut
FRGS & $N_\sigma$ \\ \midrule
$\ell=1$ & 	0.204	\\
$\ell=2$ & 	1.158	\\
$\ell=3$ & 	2.888	\\
$\ell=4$ & \bf	5.298	\\
$\ell=5$ & 	4.185	\\
$\ell=6$ & 	1.216	\\ \bottomrule
\end{tabular} 
}}$
\end{center}
\vspace*{-4mm}
\caption{The same as Figure~\ref{fig:examples}, for the Kaluza Klein
  pattern described in the text.  The table presents the FRGS for each
  level $\ell$.
  As with the oscillatory example from Figure~\ref{fig:examples}, reconstructing the signal using (in this example) $7\%$ of the wavelet coefficients involves a tradeoff between noise reduction and fidelity to the finer details of the injected signal.
  The FRGS, on the other hand, correctly identifies significant excesses in the $\ell=4$, $\ell=5$ and $\ell=3$ resolution levels of $5.3\sigma$, $4.2\sigma$, and $2.9\sigma$, respectively.}
\label{fig:kk}
\end{figure}

A more realistic oscillatory pattern could correspond to a Kaluza
Klein spectrum of resonances.  We consider a series of resonances
inspired by a warped extra dimension \cite{Randall:1999ee} for which
the first resonance appears at $m_1 \approx 320~\gev$ with a width of
$\Gamma_1\approx 18~\gev$, and subsequent masses and widths $m_i$ and
$\Gamma_i$ are given by
\begin{align}
m_i \approx \frac{x^{(1)}_i}{x^{(1)}_1} m_1 &&
\Gamma_i \approx \frac{x^{(1)}_i}{x^{(1)}_1} \Gamma_1,
\end{align}
where $x^{(1)}_i$ is the $i$th zero of the Bessel function $J_1(x)$.

This is a case where the signal is spread throughout the distribution,
and the FRGS
is useful to combine the significances from
the statistically independent wavelet coefficients of a given level.
In Fig.~\ref{fig:kk}, we show the wavelet transform of this signal on
top of the ATLAS background model.  Individual wavelet coefficients
show up to $\sim 4\sigma$ deviations from the background model at
$\ell=3$ and $\ell=4$, corresponding to the first three resonances in
the tower. Combining the significances at each level, the FRGS
indicates a $5.3\sigma$ deviation at $\ell=3$, along with
3--4$\sigma$ excesses at other resolutions.  

This example illustrates the
power of the wavelet transform and FRGS to tease out oscillatory
signals, even when the `frequency' of the signal is not constant.
Our analysis could be just as easily applied to cases with large numbers of new states, for example~\cite{clockwork} and~\cite{DAgnolo:2019cio},
and to models with multiple resonances at arbitrary masses and widths.

\subsection{ATLAS Distribution}
\label{sec:atlas}

\begin{figure}[ht]
\centering
\def\hwid{0.56}
\hspace*{-0.02\textwidth}
\includegraphics[height=\hwid\textwidth]{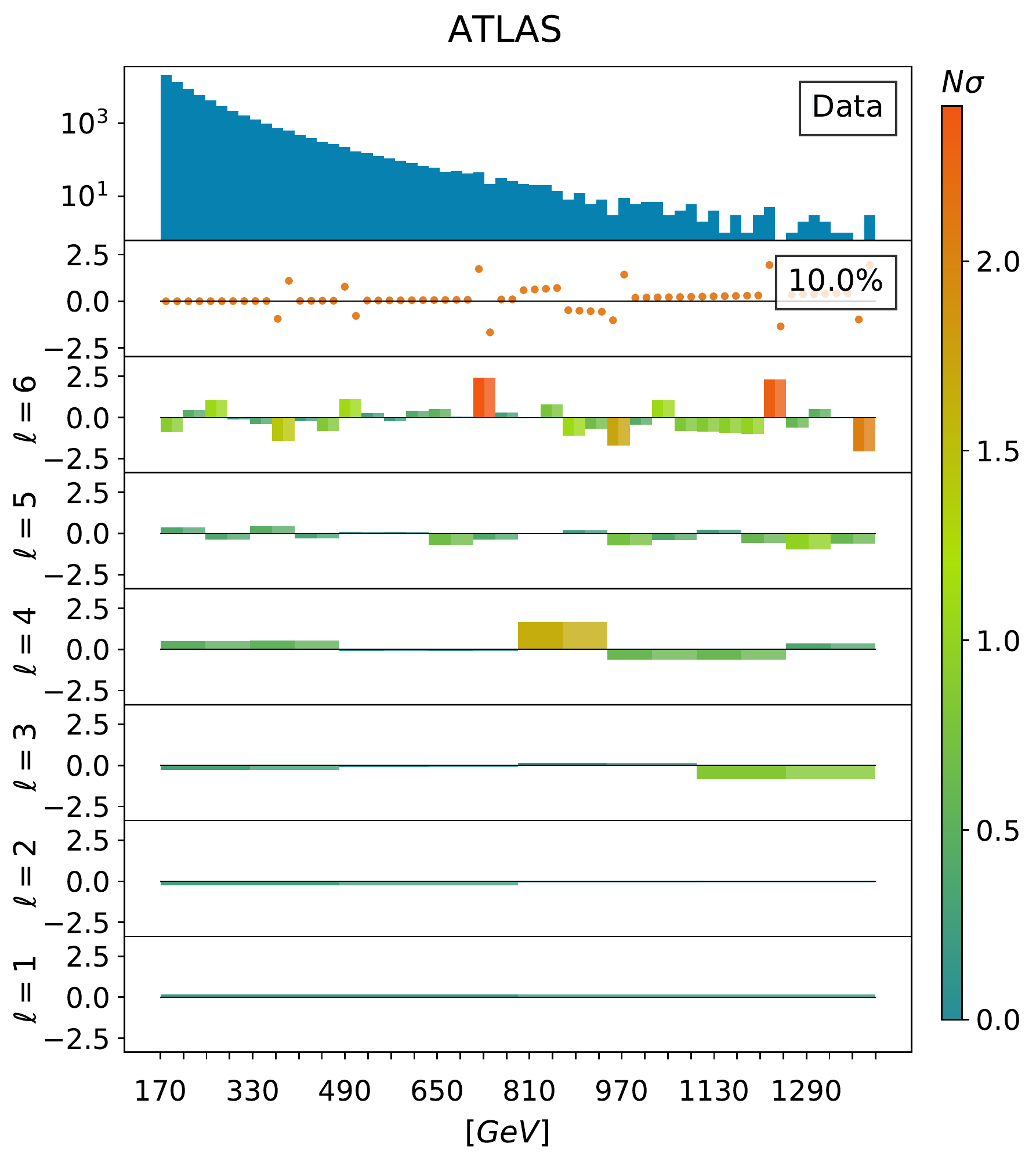}
\hspace*{-0.02\textwidth}
\includegraphics[height=\hwid\textwidth]{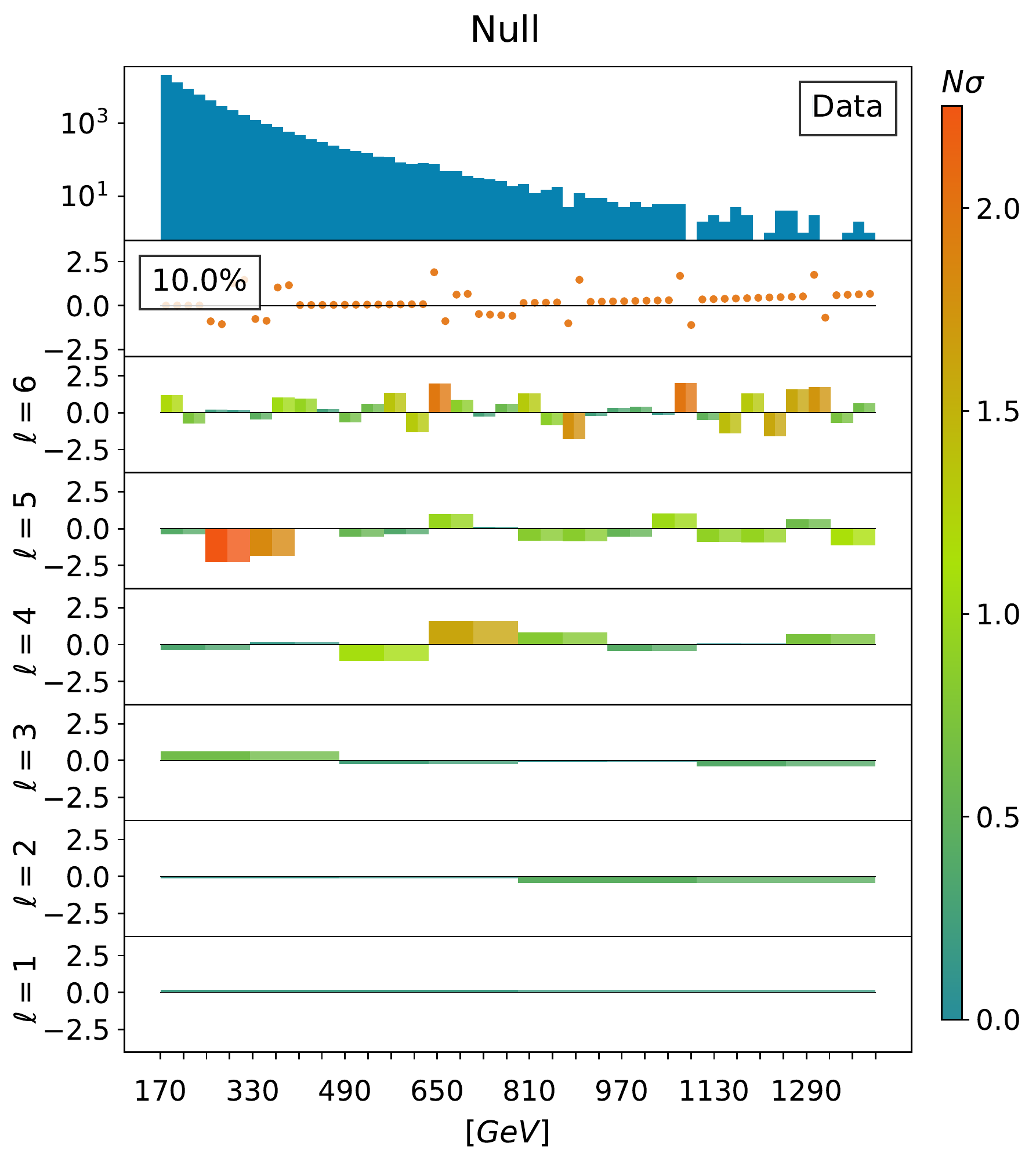} 
\begin{small} \begin{tabular}{c|llllll} 
\toprule
FRGS 	& $\ell=1$	& $\ell=2$	& $\ell=3$	& $\ell=4$	& $\ell=5$	& $\ell=6$	\\ \midrule
ATLAS $N_\sigma$ & 0.202 	& 0.0620 	& 0.0553	& 0.0831	& 0.000532	& 0.589 \\ 
Null  $N_\sigma$ & 0.0827 	& 0.0682 	& 0.0354	& 0.0990	& 0.189		& 1.006 \\ \bottomrule
\end{tabular} \end{small}
\caption{Top: wavelet analysis of the ATLAS $m_{\gamma \gamma}$ data
  (left) and an example null hypothesis distribution (right).  Bottom:
  Fixed resolution global significance at each level for the
  ATLAS and `Null' data sets.}
\label{fig:atlas_frgs}
\end{figure}


Our final example is to analyze the actual ATLAS di-photon
distribution~\cite{atlas}, shown in Fig.~\ref{fig:atlas}.  While we
already know from the original analysis that it contains no
indications of new physics, we can still use it as an example for our
wavelet analysis tool in a realistic setting.  The wavelet transform
of the ATLAS di-photon data is shown in the left pane of
Fig.~\ref{fig:atlas_frgs}.  The fluctuations in all wavelet
coefficients are small and reach the $2\sigma$ level in only two
places.  In the table below we give the FRGS at each level and, as
expected, the ATLAS distribution indicates no signs of new physics. In
fact, the wavelet coefficients appear to be slightly more consistent
with the null hypothesis than one would naively expect.  For instance,
given the 64 bins translated into 32 coefficients at level $\ell = 6$
or 64 coefficients altogether we would expect around 20 to deviate at
the $1\sigma$ level and 3 to deviate at the $2\sigma$ level.

We can compare the ATLAS result to a background-only set of toy data
based on per-bin Poisson statistics, shown in the left pane of
Fig.~\ref{fig:atlas_frgs}.  Indeed, the statistical fluctuations are
slightly more pronounced. From the corresponding Table we see that the
difference is most visible at the level $\ell=5$.
While it is beyond our ability to delve further in a meaningful way
into what the origin of this feature is, one could imagine that it is
the result of correlations between nearby $m_{\gamma \gamma}$ bins,
which our analysis treats as independent. Correlations between bins
and bin migration certainly have the potential to soften the
statistical anomaly. In fact, one could imagine that the wavelet analysis might
potentially offer a means to obtain interesting insights into such
correlations in a way that is orthogonal to traditional approaches.

\section{Outlook} 

Wavelets are a novel way to represent data in a way which, by
simultaneously retaining information on multiple scales, allows for a
flexible search for features on multiple scales.  We have applied the
Haar wavelet to a one-dimensional kinematic distribution, and
demonstrated that local features of various sizes and global
structures can both be disentangled.  As toy examples we have shown
how narrow and wide bumps, a bump-dip, and a KK-inspired oscillation
pattern can be extracted from toy data as well as from an ATLAS
di-photon mass spectrum. The background model is a simple,
model-independent fit function.\medskip

We have discussed how the different features can be separated and
understood from a universal analysis of wavelet coefficients, and how
we can perform a statistical analysis on the wavelet coefficients. In
the absence of correlations the translation from mass bins to wavelet
coefficients is a simple linear transformation without any loss of
information. Including correlations requires a proper statistical
treatment. One of the most interesting aspects of our analysis is the
fixed resolution global significance (FRGS) determined from one set of
wavelet coefficients. To visualize the relevance of an anomaly we can
also reconstruct the signal-background combination from the leading
wavelet coefficients and find very good agreement with the injected
signal. We hope that they will find fruitful use in future analysis of
LHC data.\medskip

Our Kinematic Wavelet Analysis Kit (\kwak) is available as a numerical python package 
at \texttt{https://github.com/alexxromero/kwak\_wavelets}.

\bigskip
\section*{Acknowledgments}

We acknowledge conversations with Daniel Whiteson, and inspiration from Carlos -- who we hope will
not disappoint us.
The work of BGL and TMPT is supported in part by NSF Grant No.~PHY-1620638.
The work of BGL is also supported in part by the Chair's Dissertation Fellowship from the UCI Department of Physics~\&~Astronomy.
The work of AR is supported in part by NSF Grant No.~PHY-1633631.
This work was performed in part at the Aspen Center for Physics, 
which is supported by NSF grant PHY-1607611. 

\appendix

\section{Statistical Method}
\label{appx:tool}

Our statistical analysis is conducted on the coefficients of the Haar wavelet transformation of a binned 
distribution $f$, where $f_i$ is the number of events in the $i^\text{th}$ bin of the distribution. For this 
integer-valued signal we use a wavelet transformation with $\tilde{f}_{L,1} = f_1 - f_2$, $\tilde{f}_{L-1, 1} = f_1 + f_2 - f_3 - f_4$, and so on, based on a basis of functions $h_{\ell, m}$ which are orthogonal but not normalized.

Given some hypothesis $H_0$ that predicts the mean expected value $\mu_i$ for each $f_i$ 
and under the assumption of Poisson statistics, the probability distribution  
$P(\tilde{f}_{\ell,m} | H_0)$ can be shown to have the same form as Eq.\eqref{eq:poiss_h}. 
The derivation is simple, and relies on the observation that 
every $\tilde{f}$ can be written in the form $\tilde{f} = f_a - f_b$ for some 
Poisson-distributed variables $f_a$ and $f_b$. For wavelet coefficient $\tilde{f}_{\ell,m}$, 
these $f_{a,b}$ are given by
\begin{align}
f_{a} =\;& \sum_{j_{a,\text{ min}}}^{j_{a,\text{ max}}} f_j  ,
&
j_{a,\text{ min}} =\;&  2^{L - \ell + 1}(m - 1) + 1
&
j_{a,\text{ max}} =\;& 2^{L - \ell}(2m - 1) 
\notag\\
f_{b} =\;& \sum_{j_{b,\text{ min}}}^{j_{b,\text{ max}}} f_j  ,
&
j_{b,\text{ min}} =\;&  2^{L - \ell}(2m - 1) + 1 
&
j_{b,\text{ max}} =\;& 2^{L - \ell + 1}m .
\label{effmu}
\end{align}
As $f_{a,b}$ are both sums of Poisson-distributed variables, $f_a$ and $f_b$ follow 
Poisson distributions with mean values
\begin{align}
\mu_{a,b} = \sum_{\text{min } j_{a,b}}^{\text{max } j_{a,b}}\mu_j ,
\end{align}
and $P(\tilde{f} | H_0)$ is the Skellam distribution
\begin{equation}
P( \tilde{f}_{\ell,m} = \tilde{f} | H_0) = e^{-\mu_a - \mu_b} \left( \frac{\mu_a}{\mu_b} \right)^{\tilde{f}/2} \mathcal I_{\tilde{f} }(2 \sqrt{\mu_a \mu_b}).
\label{eq:poiss_ab}
\end{equation}
%

Signals of new physics may in general be manifested in the wavelet coefficients as positive or negative
 fluctuations in $\tilde{f}$ away from the mean expected value $\mu = \mu_a - \mu_b$, and so we use a two-tailed test
  to quantify the significance of a deviation. 
Given a background hypothesis $H_0$ and the measured value $\tilde{f}$ for each wavelet coefficient, 
we define the $p$-value as the likelihood of obtaining an outcome that is at least as extreme as the measured
 value, where by ``more extreme'' we mean ``less probable''. 
Expressed in terms of the finite sum over all $i$ such that $P(i | H_0) > P(\tilde{f} | H_0)$:
\begin{equation}
1-p =  \sum_{\forall i:\; P(i | H_0) > P(\tilde{f} | H_0) } P(i | H_0).
\label{eq:1pval}
\end{equation} 
An excess can also be characterized by the number of standard deviations between $\tilde{f}$ and the 
mean expected value $\mu$, which in the Gaussian limit $\mu_a + \mu_b \gg 1$ is given by
\begin{equation}
N_\sigma = \sqrt{2} \erf^{-1} (1 - p).
\label{eq:nsigma}
\end{equation}
Even in the non-Gaussian limit of the Skellam distribution, it is often convenient to reference this definition 
of $N_\sigma(p)$ as a proxy for the $p$-value.

\paragraph{Fixed Resolution Global Significance:}
In a distribution with statistically independent bins, the wavelet coefficients within a given level $\ell$ are 
also mutually independent, making it straightforward to combine their significances.
Following~\cite{fisher}, the test statistic $q_i = -2 \ln p_i$ obeys a $\chi^2$ distribution with two degrees of 
freedom: thus, the combined test statistic $q = q_1 + q_2 + \ldots + q_k$ with $k$ independent wavelet 
coefficients follows the $\chi^2$ distribution with $2k$ degrees of freedom, $\chi^2_{2k}$.

After computing $q_\ell = \sum q_m$ from all $m = 1, 2, \ldots, 2^{\ell -1}$ coefficients in the $\ell^\text{th}$ level
of the wavelet transformation, we calculate the fixed resolution global significance from the cumulative 
distribution function of the $\chi^2_{(2^\ell)}$ distribution:
\begin{align}
D(\chi^2_{2k}) = \frac{\gamma\left(k, \frac{1}{2} \chi^2 \right)}{\Gamma(k)} 
\quad\longrightarrow\quad
p_\ell =  1 - \frac{\gamma\left(2^{\ell - 1} , \frac{1}{2} q_\ell \right) }{\Gamma(2^{\ell - 1})},
\label{eq:incgamma}
\end{align}
where $\gamma(k, z)$ is the lower incomplete gamma function. This $p_\ell$ represents the likelihood that 
Poisson sampling of the hypothesis $H_0$ would return a value for the combined test statistic that is at least 
as large as $q_\ell$.

The fixed resolution global significance is particularly powerful for identifying signals that exhibit oscillatory 
behavior, whereas well localized signals such as simple bumps and bump-dips are more likely to be 
best identified by a small set of individual wavelet coefficients.

\section{Kinematic Wavelet Analysis Kit} 
\label{appx:kwak}

The Kinematic Wavelet Analysis Kit (\kwak) is a numerical Python package for the 
statistical analysis of binned distributions of a single kinematic variable.
Its central function is to determine the probability distribution for each coefficient of the 
wavelet transformation of the data, and to identify the most significant deviations
from a given background hypothesis.
The \kwak\ package also provides a number of plotting options for displaying the results of the analysis,
and is
available online at \texttt{https://github.com/alexxromero/kwak\_wavelets},
or installed via the command
\begin{align}
\texttt{pip install kwak} \nonumber
\end{align}
for either Python~2 or Python~3.

KWAK provides multiple options for calculating the probability distribution 
for each wavelet coefficient, including an exact approach based on Eq.\eqref{eq:poiss_ab},
and three related approximate methods.

\begin{figure}[ht]
\centering
\hspace*{-0.018\textwidth}
\includegraphics[height=0.44\textwidth]{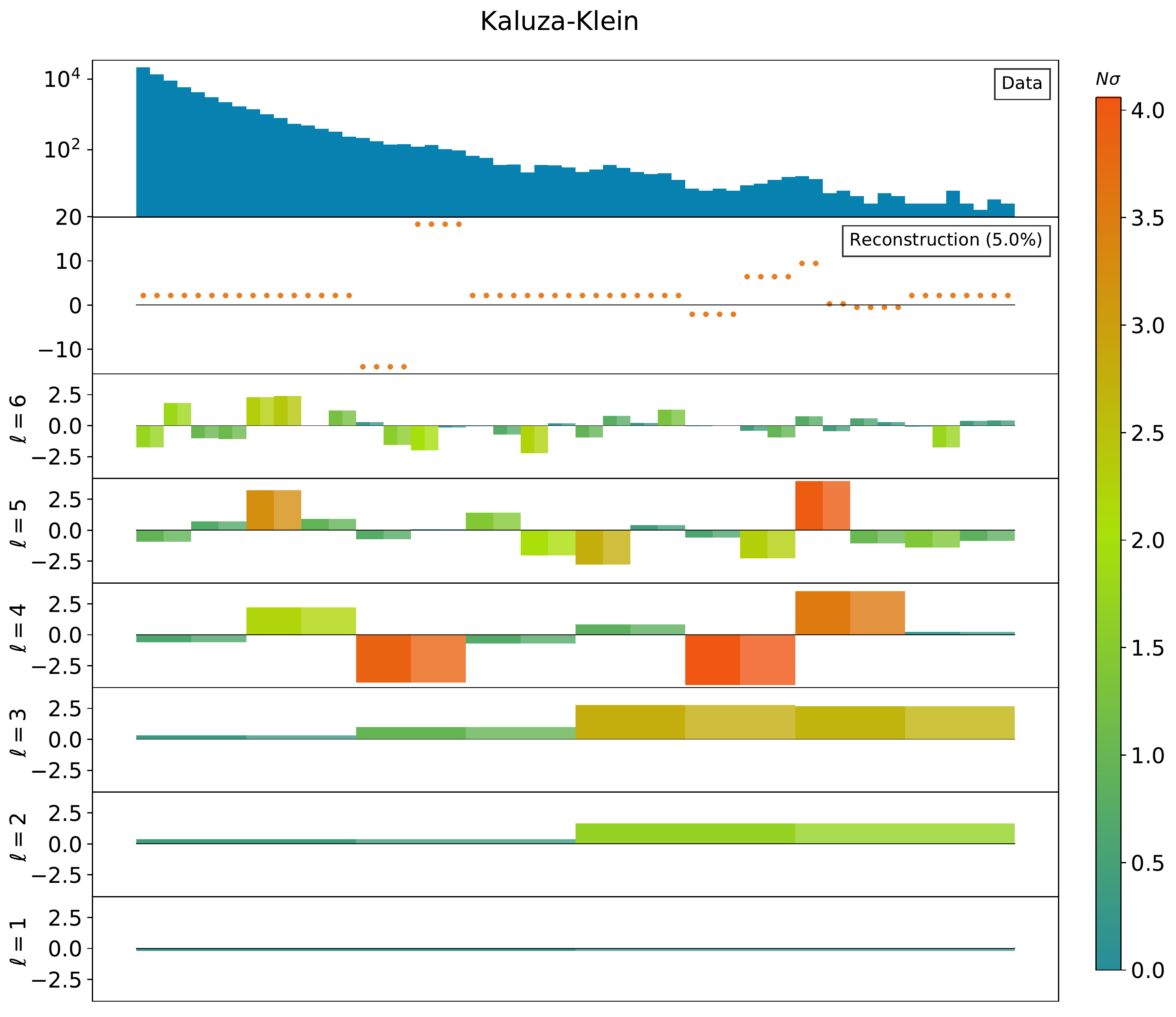}
\includegraphics[height=0.44\textwidth]{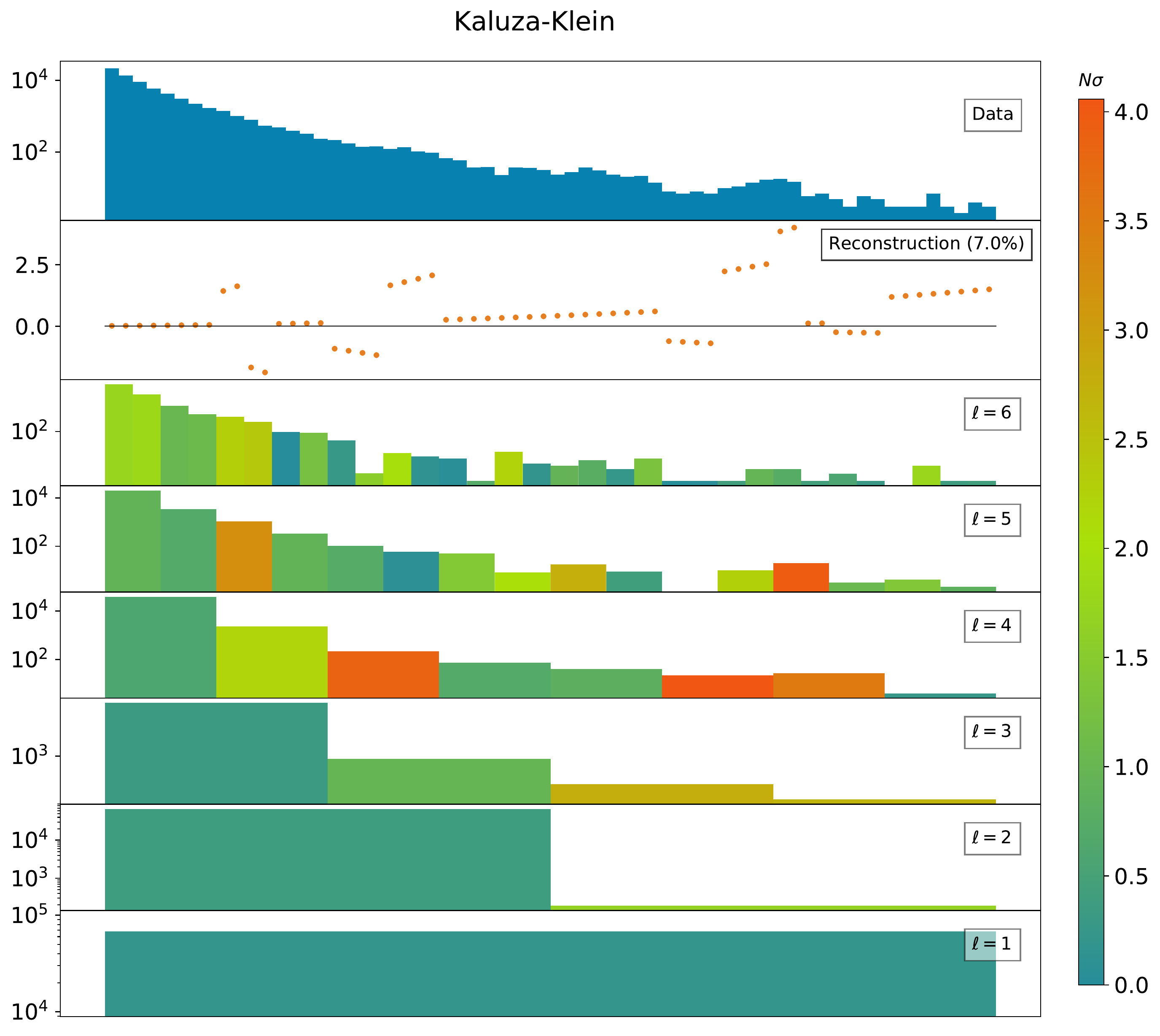} 
\caption{Left: the Kaluza-Klein model from the main text is used as a demonstration of the \texttt{nsigScalogram} plot with $\texttt{reconstruction\_scaled} = \texttt{nsigma\_colorcode} = \texttt{False}$.
Right: a \texttt{wScalogram\_nsig} plot of the same Kaluza-Klein model with 
$\texttt{reconstruction\_scaled} = \texttt{nsigma\_colorcode} = \texttt{logscale} = \texttt{True}$ 
and $\texttt{firsttrend} = \texttt{False}$.
  }
\label{fig:kk-nsig}
\end{figure}

\begin{figure}[ht]
\centering
\hspace*{-0.02\textwidth}
\includegraphics[height=0.46\textwidth]{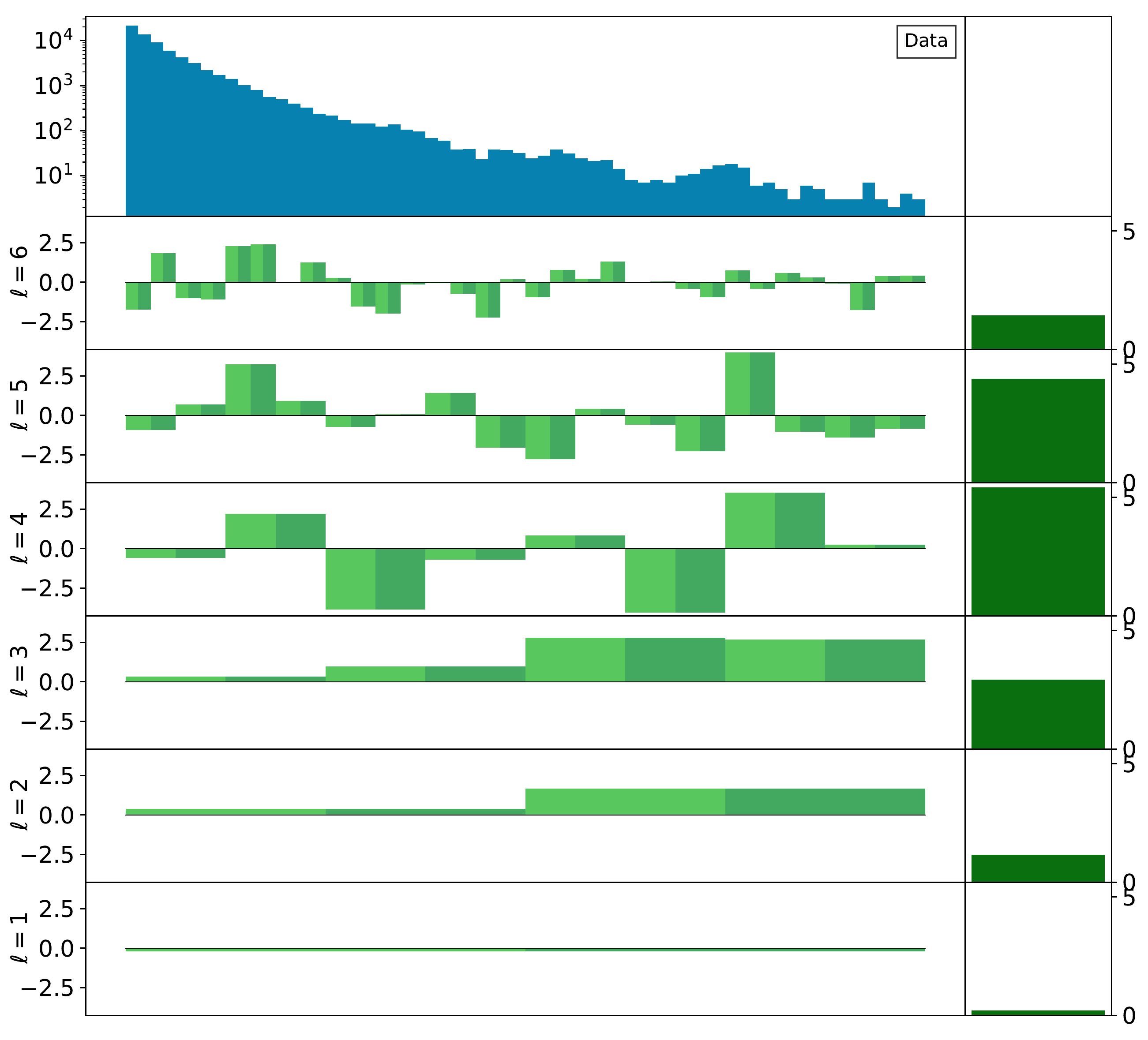}
\includegraphics[height=0.46\textwidth]{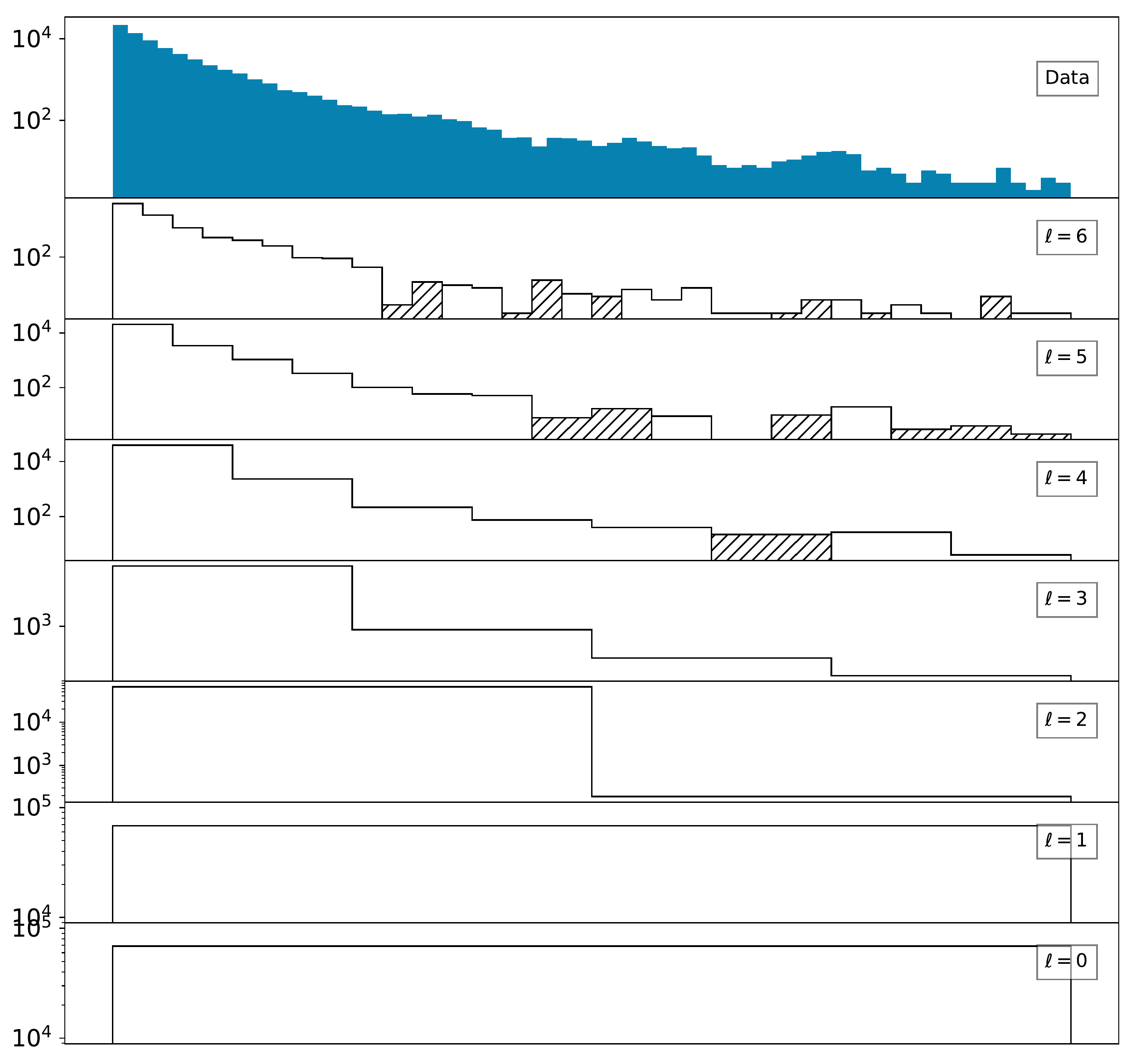} 
\vspace*{-6mm}
\caption{Using the same Kaluza-Klein model, two further plot examples are shown. 
Left: \texttt{nsigFixedRes} with $\texttt{nsigma\_colorcode} = \texttt{False}$.
Right: \texttt{wScalogram} with $\texttt{filled} = \texttt{False}$ on a logarithmic scale with $\texttt{firsttrend} = \texttt{True}$.
  }
\label{fig:kk-frgs}
\end{figure}

\paragraph{Exact Method:}

The exact approach is based on the assumption of Poisson statistics,
and is valid specifically for kinematic distributions where 
the systematic error can be neglected. In this case 
the $p$-value for every coefficient in the wavelet transformation 
can be calculated by evaluating Eq.\eqref{eq:1pval} directly, using the
Skellam distribution of Eq.\eqref{eq:poiss_ab}.

This approach can be computationally intensive: the sums over less-extreme probabilities in Eq.\eqref{eq:1pval} require repeated evaluation of the $k^\text{th}$ modified Bessel function of the first kind, where $k = \tilde{f}$ is an integer that scales with the number of events in the associated bins.
Our \kwak\ implementation uses the \texttt{mpmath} Python library to conduct the calculation at arbitrary precision, to handle the exponentially large or small values of $\mathcal I_k (z)$.
\kwak\ also uses \texttt{mpmath} to accommodate data sets with especially large fluctuations, where the individual probabilities $P(\tilde{f} | H_0)$ would otherwise be smaller than the floating point error.

These calculations are implemented in \kwak\ in the \texttt{kwak.exact} class:
\begin{itemize}
\item[] \texttt{kwak.exact(data, hypothesis, outputdir=\textit{None})} 
\end{itemize}
where \texttt{data} and \texttt{hypothesis} are one-dimensional arrays of equal length. 
If a value is provided for the optional keyword argument \texttt{outputdir}, the results of the
analysis will be saved to a newly created directory with that name.

Instantiating the \texttt{kwak.exact} class creates several objects, including:
\begin{itemize}
\item \texttt{self.Nsigma}: the $p$-value for every wavelet coefficient, mapped to a value of ``$N_\sigma$'' following Eq.\eqref{eq:nsigma}.
\item \texttt{self.NsigmaFixedRes}: the fixed resolution global significance for each level of the wavelet transformation.
\item \texttt{self.Histogram}: the probability distribution for each wavelet coefficient $P(i | H_0)$, calculated only for the values of $i$ necessary to evaluate the sum of Eq.\eqref{eq:1pval}.
\end{itemize}

Evaluating the 64-bin diphoton examples of Fig.~\ref{fig:examples} takes $\mathcal O(500)$ seconds when using the exact approach.

\paragraph{Approximate Methods:}

In situations where the precision of the exact method is unnecessary, or where the effect of systematic uncertainties cannot be neglected, it may be more appropriate to calculate $P( \tilde{f} | H_0)$ using one of the approximate methods of the \texttt{kwak.nsets} class.
These three related approaches each approximate the wavelet coefficient probability distributions by generating a large number, $N_\text{sets}$, of pseudo-random ``data'' sets drawn from the background-only hypothesis $H_0$ using Poisson statistics.\footnote{Systematic effects could in principle be mimicked by adding some smearing to the Poisson mean $\mu_i$ in each bin of the pseudodata, but such modifications are left to the user.}
After performing a wavelet transformation on each pseudodata set, the \texttt{nsets} class assembles a histogram $D_{\ell,m}(\tilde{f} | H_0)$ for each wavelet coefficient, counting the number of pseudoexperiments $D_{\ell,m}$ which return a value $\tilde{f}_{\ell,m} = \tilde{f}$ for the $(\ell, m)$th wavelet coefficient.
The probability distribution for that coefficient is approximated by: 
\begin{equation}
P_{\ell,m}(\tilde{f} | H_0) = \frac{D_{\ell,m}(\tilde{f} | H_0)}{N_\text{sets}},
\label{eq:phistogram}
\end{equation}
where the histogram $D_{\ell, m}$ includes the values from $(N_\text{sets} -1)$ pseudoexperiments as well as the real data.
Our choice to use an unnormalized wavelet transformation ensures that $\tilde{f} = \mu_1 - \mu_2$ is integer-valued.

This approach is limited by the fact that Eq.\eqref{eq:phistogram} does not resolve any probabilities smaller than $P_\text{min} = N_\text{sets}^{-1}$. 
Reliably distinguishing $4\sigma$ from $5\sigma$ deviations, for example, requires somewhat better than $N_\text{sets} = 10^7$, after accounting for the fact that there may be several values of $\tilde{f}$ for which $D(\tilde{f} | H_0) = 1$. 
Nevertheless, relatively small $N_\text{sets}$ can be sufficient for identifying deviations in the data, in much less time than is possible with \texttt{exact}.
It also handles non-Gaussian distributions well: no assumptions about the shape of 
$P_{\ell,m}(\tilde{f} | H_0)$ are built in to this analysis.

The default implementation of the \texttt{nsets} method described above can be expanded with one of the two following options:
\begin{itemize}
\item \texttt{fastGaussian}: calculates the mean and standard deviation for each histogram $D_{\ell,m}$
\item \texttt{extrapolate}: applies a functional fit to the histogram $D_{\ell,m}$, using an approximation of the Skellam distribution
\end{itemize}
With the first option, rather than defining the probability distribution $P_{\ell, m}$ and the $p$-value $p_{\ell, m}$, $N_\sigma$ is calculated directly and very simply from the mean $\mu(\tilde{f})$ and standard deviation $\sigma(\tilde{f})$ of the histogram $D_{\ell, m}$:
\begin{equation}
N_\sigma(\tilde{f}_{\ell, m} ) = \frac{ \tilde{f}_{\ell, m} - \mu(\tilde{f}_{\ell, m}) }{\sigma(\tilde{f}_{\ell, m}) }.
\end{equation}
In the Gaussian limit of the Skellam distribution, $\mu_1 + \mu_2 \gg 1$, the \texttt{fastGaussian} approach provides a much better approximation of $N_\sigma$ for large fluctuations,
\begin{equation}
N_\sigma > \sqrt{2} \erf^{-1} \left(1 - (\text{few}) \times N_\text{sets}^{-1} \right),
\end{equation}
compared to what is possible with the default \texttt{nsets} method.

However, as seen in the left panel of Fig.~\ref{fig:distris}, when $\mu_1 + \mu_2 < 1$ the Skellam distribution does not resemble a Gaussian at all, instead peaking sharply at $\tilde{f} = 0$. 
For rare processes with small but well-understood backgrounds, one or two events in some region of a kinematic distribution may be highly significant,
requiring us to employ a better approximation of the Skellam distribution.

The \texttt{extrapolate} option is designed to handle both limits smoothly. It uses the curve fitter from \texttt{scipy.optimize} to fit the histograms $D_{\ell, m}$ with a modified Gaussian function
\begin{equation}
D_{\ell, m}(\tilde{f} ) \approx n \exp\left( - \frac{1}{2} \left( \frac{\tilde{f} - \mu}{\sigma} \right)^2 - \gamma \left|\tilde{f} \right|^p \right) 
\label{eq:fitfunction}
\end{equation}
for some $p \approx 1$ and $\gamma \geq 0$. 

Unlike the default version of \texttt{nsets} or the \texttt{fastGaussian} alternative, 
the \texttt{extrapolate} option requires a relatively large minimum value of $N_\text{sets}$
in order to run smoothly. 
If $N_\text{sets}$ is not large enough to generate nonzero entries in the histogram $D(\tilde{f})$ 
beyond the central values of $\tilde{f} = 0, \pm 1, \pm2$, 
then the five parameter fit of Eq.\eqref{eq:fitfunction} might not have a well-defined best fit point.
For bins in the kinematic distribution with expected mean values $\mu_i \lesssim 10^{-1}$, it may be necessary to use $N_\text{sets} > 10^5$ to guarantee that \texttt{extrapolate} will provide a good fit for the probability distribution.
\medskip

All three approximate methods are integrated into the \texttt{nsets} class:
\begin{itemize}
\item[] \texttt{kwak.nsets(data, hypothesis, nsets, seed=\textit{int}, outputdir=\textit{None},\\ {\color{white} kwak.nsets} fastGaussian=\textit{Boolean}, extrapolate=\textit{Boolean})} 
\end{itemize}
where $\texttt{nsets} = N_\text{sets}$ determines the number of pseudoexperiments to generate, and \texttt{seed} specifies the seed to be used for the random number generator.
By default, \texttt{fastGaussian} and \texttt{extrapolate} are set to \texttt{False}. 
Given conflicting inputs $\texttt{fastGaussian}=\texttt{True}$ and $\texttt{extrapolate}=\texttt{True}$, the $\texttt{fastGaussian}=\texttt{True}$ option takes precedence,
and the \texttt{extrapolate} calculation will not be performed.

The \texttt{nsets} class also has \texttt{self.Nsigma}, \texttt{self.NsigmaFixedRes}, and \texttt{self.Histogram} objects; the only difference from the \texttt{exact} class is that for \texttt{nsets} the \texttt{self.Histogram} is the collection of histograms $D_{\ell, m}$, rather than the probability distributions $P_{\ell, m} = D_{\ell, m} \times N_\text{sets}^{-1}$.

\paragraph{Comparison:}

A rough guide to when (and when not) to use each of the four methods is given below:
\begin{itemize}
\item \texttt{exact}: Valid whenever the systematic uncertainties can be neglected. Especially useful at quantifying large fluctuations, and for cases where the evaluation time is not important.
\item \texttt{nsets} (default): Provides fast analysis, best suited for data sets with moderate or small fluctuations. Valid for non-Gaussian probability distributions.
\item \texttt{fastGaussian}: As fast as the default \texttt{nsets}, and able to distinguish between moderate and large fluctuations. Only valid for kinematic distributions where multiple events are expected in every bin.
\item \texttt{extrapolate}: Expands the default \texttt{nsets} method to distinguish between moderate and large fluctuations, even in the non-Gaussian limit. Requires a larger minimum $N_\text{sets} \sim 10^5$ when operating in this limit.
\end{itemize}
As both the default \texttt{nsets} and the \texttt{fastGaussian} approximations can be run with $N_\text{sets} = 10^3 \text{ -- } 10^4$, these methods are the best choices if the analysis must be repeated many times.

The \texttt{fastGaussian} method remains accurate even for small values of $N_\text{sets}$: for example, calculating the FRGS for the Kaluza-Klein model shown in Fig.~\ref{fig:kk} with $N_\text{sets} = 10^3$ gives:
\begin{center}
\begin{tabular}{r|cccccc} \toprule
KK FRGS ($N_\sigma$)	& $\ell=1$	& $\ell=2$	& $\ell=3$	& $\ell=4$	& $\ell=5$	& $\ell=6$	\\ \midrule
\texttt{exact}: 			& 0.204	& 1.158	& 2.888	& 5.298	& 4.185 	& 1.216	\\
\texttt{nsets-default}:		& 0.422	& 0.850	& 2.422	& 3.022	& 2.959	& 0.893	\\
\texttt{nsets-fastGaussian}:& 0.230	& 1.157	& 2.859	& 5.267	& 4.459	& 1.497	\\ \bottomrule
\end{tabular}
\end{center}
Considering that \texttt{fastGaussian} with $N_\text{sets} = 10^3$ already approaches the accuracy of the \texttt{exact} method, and evaluates almost 1000 times more quickly, there is a real benefit to taking the Gaussian approximation if appropriate.
\medskip

In the Gaussian limit with multiple events expected in every bin, the \texttt{extrapolate} approach can be used with a smaller minimum $N_\text{sets} \ll 10^5$. 
Below $N_\text{sets} < 10^4$, the evaluation time becomes dominated by the curve fitting function, so that $N_\text{sets} = 10^3$ takes as long to evaluate as $N_\text{sets} = 10^4$.
Thus, the primary purpose of \texttt{extrapolate} is to provide improved accuracy in the $10^4 < N_\text{sets} < 10^6$ range, especially for cases when the Gaussian approximation is not necessarily appropriate.

Around $N_\text{sets} = 1.5 \times 10^6$, the three approximate calculations and the \texttt{exact} method take equivalent amounts of time to evaluate. 
Unless systematic uncertainties are being included in the calculation, there is no benefit to running any of the \texttt{nsets} approximations with $N_\text{sets} > 10^6$, as \texttt{exact} becomes faster at this point.

\paragraph{Plotting Functions and Options:} 

The plots of Figures~\ref{fig:examples},~\ref{fig:kk}, and~\ref{fig:atlas_frgs}
are generated using one of the plot types included in the \kwak\ package, \texttt{kwak.nsigScalogram}:
\begin{itemize}
\item[] \texttt{kwak.nsigScalogram(data, hypothesis, nsigma, \textit{*kwargs})}
\end{itemize}
where \texttt{nsigma} should be the \texttt{self.Nsigma} object from an \texttt{exact} or \texttt{nsets} class.
The top two panels of this plot show a histogram of the data, and a reconstruction of the putative signal  
using only the wavelet coefficients with the largest deviations away from the background hypothesis.
The remaining panels show the value of $N_\sigma$ for each wavelet coefficient.

In addition to the mandatory arguments, a number of optional keyword arguments 
can be used to change characteristics of the plot:
\begin{itemize}
\item For the reconstruction of the signal:
	\begin{itemize}
	\item $\texttt{nsigma\_min}=x$: Uses only wavelet coefficients with $N_\sigma > x$.
	\item $\texttt{nsigma\_percent}=x$: Uses only the most significant $x\times 100\%$ wavelet coefficients.
	\item $\texttt{reconstruction\_scaled} = \textit{Boolean}$: Provides an option to divide all of the entries in the reconstructed signal by the square root of the mean expected value for that bin, so that the $y$ axis corresponds loosely to ``$N_\sigma$'' rather than the number of events in the signal.
	\end{itemize}
\item $\texttt{nsigma\_colorcode} = \textit{Boolean}$: Color codes the plot of the wavelet coefficients with a scheme based on the size of $N_\sigma$.
\item $\texttt{title} = \textit{str}$: Prints a title above the plot, in size 18 font.
\item $\texttt{xlabel} = \textit{str}$: Prints a label for the $x$ axis, in size 14 font.
\item $\texttt{outputfile} = \textit{str}$: Saves the plot as a PNG file with name \texttt{"outputfile"}.
\end{itemize}
As an example of the default output of \texttt{nsigScalogram},
Fig.~\ref{fig:kk-nsig} shows the Kaluza-Klein model of Fig.~\ref{fig:kk}
but with $\texttt{reconstruction\_scaled} = \texttt{nsigma\_colorcode} = \texttt{False}$.

Rather than plotting $N_\sigma$ for each wavelet coefficient, the plotting function \texttt{kwak.wScalogram\_nsig} replaces $N_\sigma$ with the values of the wavelet coefficients themselves.
In addition to the keyword arguments available for \texttt{nsigScalogram}, \texttt{kwak.wScalogram\_nsig} has an option to plot the values of the wavelet coefficients on a logarithmic scale:
\begin{itemize}
\item $\texttt{logscale} = \textit{Boolean}$.
\end{itemize}
Negatively signed wavelet coefficients are shown as positive values with hatched lines on the logarithmic plot,  
as shown in the right panel of Fig.~\ref{fig:kk-nsig}.
A second additional optional argument, $\texttt{firsttrend} = \textit{Boolean}$, determines whether or not the value of the $\tilde{f}_{\ell=0}$ coefficient is shown.

In the plots of the main text, the FRGS is typically shown as a separate table. Another plotting method, \texttt{kwak.nsigFixedRes}, shows the FRGS $N_\sigma$ value as an additional column on the right:
\begin{itemize}
\item[] \texttt{kwak.nsigFixedRes(data, hypothesis, nsigma, nsigma\_FRGS, \textit{*kwargs})}
\end{itemize}
also with the optional keyword arguments corresponding to color-coding and plot labels.
An example with the default color coding is shown in the left panel of Fig.~\ref{fig:kk-frgs}.

Finally, to display the wavelet transformation of the data without any reference to the statistical analysis, we provide
\begin{itemize}
\item[] \texttt{kwak.wScalogram(data, \textit{*kwargs})}
	\begin{itemize}
	\item $\texttt{logscale} = \textit{Boolean}$
	\item $\texttt{firsttrend} = \textit{Boolean}$
	\item $\texttt{filled} = \textit{Boolean}$
	\item $\texttt{outputdir} = \textit{str}$
	\end{itemize}
\end{itemize}
where the new optional argument \texttt{filled} determines whether or not to fill the histograms for the wavelet coefficients with a solid color. 
As before, negative coefficients on the logarithmic scale are shaded with hatch marks. An example with $\texttt{filled} = \texttt{False}$
is shown in the right panel of Fig.~\ref{fig:kk-frgs}.

\medskip

For additional control over the relative sizes of the individual panels in each plot, the range of $y$ values shown for a particular panel, the text displayed inside the legends, or other similar details, 
the user can edit the relevant parameters directly in \texttt{nsigmaplots.py} and \texttt{scalograms.py} in
the \texttt{kwak/plotting} folder.



\end{document}